\documentstyle[12pt]{article}

\newcommand{\eps}{\varepsilon}
\newcommand{\e}{{\sf e}}

\def\N{{\rm I\kern-.1567em N}}

\def\R{{\rm I\kern-.1567em R}}
\def\C{{\rm C\kern-4.7pt
\vrule height 7.7pt width 0.4pt depth -0.5pt \phantom {.}}}
\def\Z{{\sf Z\kern-4.5pt Z}}

\begin{document}

\newtheorem{theorem}{Theorem}[section]
\renewcommand{\thetheorem}{\arabic{section}.\arabic{theorem}}
\newtheorem{definition}[theorem]{Definition}
\newtheorem{deflem}[theorem]{Definition and Lemma}
\newtheorem{lemma}[theorem]{Lemma}
\newtheorem{example}[theorem]{Example}
\newtheorem{remark}[theorem]{Remark}
\newtheorem{remarks}[theorem]{Remarks}
\newtheorem{cor}[theorem]{Corollary}
\newtheorem{pro}[theorem]{Proposition}
\newtheorem{proposition}[theorem]{Proposition}

\renewcommand{\theequation}{\thesection.\arabic{equation}}

\begin{titlepage}

\begin{center}
{\Large\bf Radiation Reaction and Center Manifolds}\\
\vspace{1cm}
{\large Markus Kunze}\medskip\\
        Mathematisches Institut der Universit\"at K\"oln\\
        Weyertal 86, D-50931 K\"oln, Germany\\
        email: mkunze@mi.uni-koeln.de\bigskip\\
{\large Herbert Spohn}\medskip\\
        Zentrum Mathematik and Physik Department, TU M\"unchen\\
        D-80290 M\"unchen, Germany\\
        email: spohn@mathematik.tu-muenchen.de\bigskip\\
\end{center}
\date{Dec. 30, 1998}
\vspace{1cm}

\begin{abstract}
\noindent We study the effective dynamics of a mechanical particle
coupled to a wave field and subject to the slowly varying potential
$V(\eps q)$ with $\eps$ small. To lowest order in $\eps$ the motion
of the particle is governed by an effective Hamiltonian. In the
next order one obtains ``dissipative'' terms which describe the
radiation reaction. We establish that this dissipative dynamics has a 
center
manifold which is repulsive in the normal direction and which is global,
in the sense that for given data and sufficiently small $\eps$ the
solution stays on the center manifold forever. We prove that the 
solution of
the full system is well approximated by the effective  dissipative
dynamics on its center manifold.
\end{abstract}

\end{titlepage}


\section{Introduction}
\setcounter{equation}{0}

At the beginning of this century, in the context of the
Maxwell-Lorentz equations, radiation reaction was one of the most
outstanding problems in theoretical physics. It was left sort of
unfinished when theoreticians turned to quantum electrodynamics.
In this paper we study radiation reaction in the mathematically somewhat
more accessible case of a scalar wave field. We believe that our
results provide good indications on the effective dynamics for a
charge coupled to the Maxwell field \cite{S}.

To explain in more detail the physical context we have to
set up the model first. We consider a particle,
position $q(t)\in\R^3$ and momentum
$p(t)\in\R^3$, with ``charge'' distribution $\rho$ of total charge
\[ \e = \int d^{3}x\rho(x) \neq 0. \]
We require that $\rho$ is smooth, radial, and supported
in a ball of radius $R_{\rho}$,
$$\rho\in C_0^\infty(\R^3)\,,\quad \rho(x)=\rho_r(|x|)\,,\quad
\rho(x)=0\,\,\,\,\,\mbox{for}\,\,\,\,\,|x|\ge R_{\rho}. \eqno{(C)} $$
The particle is coupled to the scalar wave field $\phi(x,t)$
with the canonically conjugate momentum field $\pi(x,t)$, $x \in \R^3$.
In addition the particle is subject to an external potential, $V$, whose
properties will be listed below. We assume that the potential
is {\it slowly varying} on the scale of the charge distribution,
i.e., on the scale set by $R_\rho$. Formally we introduce
the dimensionless parameter $\eps$, $\eps\ll 1$, and consider
the scale of potentials $V(\eps q)$, $\eps\to 0$.
The equations of motion for the coupled system are
\begin{eqnarray}\label{system}
   \begin{array}{ll}
   \dot{\phi}(x,t)=\pi (x,t), & \dot{\pi}(x,t)=\Delta \phi (x,t)
   -\rho (x-q(t)), \\  &  \\
   \dot{q}(t)=\displaystyle\frac{p(t)}{\sqrt{1+p(t)^{2}}}, & \dot{p}(t)
   =-\eps \nabla V(\eps q(t))
   +\displaystyle\int d^3x\,\phi (x,t)\nabla \rho (x-q(t))\, .
\end{array}
\end{eqnarray}

The dynamics governed by (\ref{system}) has three distinct time scales,
well-separated as $\eps\to 0$. On the {\it microscopic} time scale,
$t = {\cal O}(1)$, the particle moves along an essentially straight line
and the field adjusts itself stationarily.
On a time scale ${\cal O}(\eps^{-1})$,
that we call the {\it macroscopic} scale, the particle feels the potential
and responds to it with an effective kinetic energy which incorporates
the coupling to the field. This scale was studied in \cite{KKS}.
The particle looses energy through radiation at a rate roughly
proportional to $\ddot{q}(t)^{2}$. Thus on the macroscopic time scale,
friction through radiation is of order $\eps$. To resolve such an effect
we have to go to even longer times or to look with higher
precision. The {\it friction} time scale is the subject of our paper.

The dynamics of (\ref{system}) is of Hamiltonian form. We need a 
few facts in the case the external potential vanishes, $V=0$. 
Then (\ref{system}) has the energy 
\[ {\cal H}_0 (\phi, \pi, q, p) = (1+p^2)^{1/2} +
   \frac{1}{2} \int d^3 x\,({|\pi(x)|}^2 + {|\nabla \phi(x)|}^2)
   + \int d^3x\,\phi(x)\rho(x-q) \]
and the conserved total momentum
\[ {\cal P} (\phi, \pi, q, p) = p + \int d^3x\,\phi(x)\nabla \pi(x). \]
The minimum of ${\cal H}_0$, at fixed ${\cal P}$, is attained at
\begin{eqnarray}
   S_{q,v} = (\phi_v (x-q), \pi_v(x-q), q, p_v)
\end{eqnarray}
where $v\in {\cal V}=\{v: |v|<1\}$, $p_v=v/\sqrt{1-v^2}$,
$\pi_v=-v\cdot\nabla\phi_v$, and $\hat\phi_v(k)
=-\hat\rho(k)/[k^2-(v\cdot k)^2]$; the
hat denotes Fourier transform. We call $S_{q,v}$
the soliton centered at $q, v$. It has the normalized energy
\begin{eqnarray*}
   {\cal E}_s(v) & = & {\cal H}_0 (S_{q,v})-{\cal H}_0 (S_{q,0}) \\
   & = & (1-v^2)^{-1/2}-1+3m_e\,\bigg[\,\frac{2-v^2}{2(1-v^2)}
   - \frac{1}{2|v|}\log\frac{1+|v|}{1-|v|}\,\bigg]
\end{eqnarray*}   
and the total momentum
\begin{eqnarray}\label{P-def}
   {\cal P}_s(v) & = & {\cal P} (S_{q,v}) \nonumber\\
   & = & v(1-v^2)^{-1/2}
   + 3m_e v\,\bigg[\,\frac{1}{2v^2 (1-v^2)} - \frac{1}{4|v|^3}
   \log\frac{1+|v|}{1-|v|}\,\bigg].\quad\quad
\end{eqnarray}
Here $m_e=\frac{1}{3}\int d^3 k\,{|\hat \rho(k)|}^2 k^{-2}$ is the
mass of the particle due to the coupling to the field.
We note that because of the Hamiltonian structure we have the identity
$v (d{\cal P}_s/dv) = (d{\cal E}_s/dv)$. It is shown in \cite{KS1}
that the map $v \mapsto {\cal P}_s (v)$ is invertible from
${\cal V}$ to $\R^3$, with inverse $P\mapsto v(P)$,
and thus $E(P):={\cal E}_s(v(P))$ is well defined.

Taking $S_{q,v}$ as initial conditions for (\ref{system}) with $V=0$ we
obtain a solution travelling at constant velocity $v$,
\[ S_{q, v}(t)=(\phi_v(x-q-vt), \pi_v(x-q-vt), q+vt, p_v)
   \,,\quad v\in {\cal V}. \]
Let us call $\{S_{q,v}: q \in\R^3, v\in {\cal V}\}$ the
six--dimensional soliton manifold, ${\cal S}$. Thus, for $V=0$,
if we start initially on ${\cal S}$ the solution remains on ${\cal S}$
and moves along the straight line $t \mapsto q^0+v^0 t$. In fact,
if we start close to ${\cal S}$, then ${\cal S}$ is approached
asymptotically, \cite {KS1}. When the particle is subject
to a slowly varying external potential, then the rough picture is
that the solution will remain close to ${\cal S}$ in the course of time.
For simplicity we assume throughout that the initial datum for
(\ref{system}) lies exactly on ${\cal S}$, i.e.,
\begin{equation}\label{ini-cond}
   (\phi(0), \pi(0), q(0), p(0))=S_{q^0,v^0}\, ,
\end{equation}
possible generalizations being discussed below.

At this point it is instructive to transform (\ref{system}) to the
macroscopic space-time scale in such a way that the field energy
remains constant. Then the macroscopic variables, denoted by a $^\prime$,
are
\begin{eqnarray*}
   & t=\eps^{-1} t^\prime\,,\quad q=\eps^{-1} q^\prime\,,\quad
   x=\eps^{-1} x^\prime\,,\quad q(t)=\eps^{-1}q'(t')\,, & \\
   & \mbox{and}\quad\phi(x,t)=\sqrt{\eps}
   \phi^\prime(x^\prime, t^\prime). &
\end{eqnarray*}   
We also set
\[ \rho_\eps(x) = \eps^{-3} \rho(\eps^{-1}x)\, . \]
In particular, $\rho_\eps(x) = 0$ for $|x|\ge\eps R_\rho$ and
$\int d^3x\rho_\eps(x)=\int d^3x\rho(x) $. With this convention,
omitting the primes and indicating explicitly the $\eps$-dependence of
$q'(t')$, we arrive at
\begin{eqnarray}\label{111}
   \ddot\phi(x,t) & = & \Delta \phi(x,t) -
   \sqrt{\eps} \rho_\eps (x-q^\eps(t)),  \\
   \dot{q}^{\eps}(t) & = & v^{\eps}(t), \nonumber\\
   m_{0}(v^{\eps}(t))\dot{v}^{\eps}(t)
   & = & -\nabla V(q(t))+\sqrt{\eps} \int d^3x\,\phi(x,t)
   \nabla\rho_\eps(x-q^\eps(t)). \nonumber
\end{eqnarray}
Here $m_{0}(v)$ is the  $3\times 3$ matrix defined through
$m_0 (v)\dot{v}=\gamma \dot{v}+\gamma^{3}(v\cdot\dot{v})v$
with $\gamma(v) = 1/\sqrt{1-v^{2}}$. Rather than momenta as
in (\ref{system}), we use velocities which turns out to be
more convenient in our context. The initial soliton (\ref{ini-cond})
transforms to
\begin{equation}\label{macro-ini}
   S^{\eps}_{q^{0},v^{0}}  = (\phi^{\eps}_{v^{0}}(x - q^{0}),
   \pi^{\eps}_{v^{0}}(x - q^{0}),q^{0},v^{0}),
\end{equation}
where $\hat\phi^{\eps}_v(k)=
-\sqrt{\eps}\hat\rho(\eps k)/[k^2-(v\cdot k)^2]$ and
$\pi^{\eps}_v=-v\cdot\nabla\phi^{\eps}_v$.
Thus, on the macroscopic scale, the total charge is
$\sqrt{\eps}\int d^3x\rho(x)$, whereas
\[ m_e = \frac{1}{3}\eps\int d^3k {|\hat\rho_\eps(k)|}^2 k^{-2}\]
is independent of $\eps$. Eqs.~(\ref{111}) are again of Hamiltonian form.
The energy
\begin{eqnarray}\label{Hmac}
   {\cal H}_{mac} (\phi, \pi, q, v) & = & \gamma(v) + V(q) +
   \frac{1}{2} \int d^3 x\,({|\pi(x)|}^2 + {|\nabla \phi(x)|}^2)
   \nonumber\\ & & + \sqrt{\eps}\,\int d^3x\,\phi(x)\rho_\eps(x-q)
\end{eqnarray}
is conserved under (\ref{111}). It is bounded from below,
as ${\cal H}_{mac}(\phi, \pi, q, v)\ge V(q)-3m_e$
independently of $\eps$.

There is another, very instructive way to think about the initial
value problem $(\ref{111}), (\ref{macro-ini})$. We prescribe
initial data at $t=-\tau$, $\tau>0$, which have finite energy and 
some smoothness.
We refer to \cite{KS1} for the precise conditions.
We solve (\ref{111}) for $V=0$ up to time $t=0$.
Then in the limit $\tau\to\infty$ the data at $t=0$
are exactly of the form (\ref{macro-ini}). For $t>0$ the external forces
are acting. Clearly this causes some mismatch, which is reflected
by a non--smoothness of the fields $(\phi, \pi)$ at the light cone
$\{x: |x| = t, t>0\}$ in the limit $\eps\to 0$.

Under suitable assumptions on $V$ and for ${|\rho|}_{L^2}$
sufficiently small we proved in \cite{KKS} that
\begin{equation}\label{v-bounds}
   |\dot{q}^\eps(t)|\le\bar{v}<1\,,\quad
   |\ddot{q}^{\,\eps}(t)| \le C\,,\quad\mbox{and}\quad
   |\stackrel{...}{q}^{\,\eps}(t)| \le C
\end{equation}
uniformly in $\eps$ and $t\in\R$, and that the limit
\begin{equation}\label{li}
   \lim_{\eps\to 0^+} q^\eps(t) = r(t)
\end{equation}
exists. Here $r(t)$ is the solution of Hamilton's equations of motion
with the effective Hamiltonian $E(p) + V(q)$, cf.~the definition
of $E(p)$ below (\ref{P-def}), which in terms of velocities read
\begin{equation}\label{vel}
   \dot{r} = u\,,\quad m(u)\dot{u} = - \nabla V(r),
\end{equation}
with initial data $r(0) = q^{0}$, $u(0) = v^{0}$.
Here $m(u)=m_{0}(u)+m_{f}(u)$, where $m_{f}(u)$ is the additional
``mass'' due to the coupling to the field defined by
\begin{equation}\label{mf-def}
   m_f (u)\dot{u} = 3m_{e}\Big(\varphi(|u|) \dot{u} +
   |u|^{-1}\varphi'(|u|)(u\cdot\dot{u})u\Big)
\end{equation}
as a $3\times 3$ matrix, where $\varphi(|v|)$ is the function appearing
in the square brackets of Eq.~(\ref{P-def}). Note that the energy
\begin{equation}\label{ener}
   H(r,u) = {\cal E}_s(u) + V(r)
\end{equation}
is conserved by the solutions to (\ref{vel}).

With this background information let us return
to the radiation reaction as  discussed by Abraham,
Lorentz, Schott, and Dirac, cf.~\cite{yag} for an excellent account.
Of course, these theoretical physicists were interested in the 
electrodynamics
of moving charges. We take here the liberty to transcribe their
arguments to the case of a scalar wave equation. For the sake of
discussion we reintroduce the bare mass $m_0$ and state the
equations for small velocities only. In our proof below, however,
we will handle all $v\in {\cal V}$.

At the beginning of this century the hope was to define a structureless
elementary charge through a point charge limit. For this program,
one had to model the charge distribution phenomenologically
with the understanding that finer details should become irrelevant
in the limit. In (\ref{111}) we adopted the Abraham model
of a rigid charge distribution. The point charge limit then corresponds
to taking in (\ref{111}) the charge distribution $\rho_\eps$
instead of $\sqrt{\eps} \rho_\eps$.
With this choice $\e=\int d^3 x\rho_\eps(x)= \int d^3 x\rho(x)$, 
whereas the
electromagnetic mass equals $\frac{1}{3}\int d^3k {|\hat\rho_\eps(k)|}^2
k^{-2}=\eps^{-1} m_e$. A formal Taylor expansion leads
to the effective equation of motion
\begin{equation}\label{portu}
   m_0 \ddot r = - \nabla V(r) - \eps^{-1} m_e \ddot{r} +
   a \e^{2} \stackrel{...}{r},
\end{equation}
valid for small velocities $\dot{r}$, with some constant $a>0$.
Eq.~(\ref{portu}) is the nonrelativistic limit of the Lorentz-Dirac
equation, \cite{rohr}. The standard argument, reproduced in many
textbooks, e.g.~\cite{jack}, (with the notable exception of
Landau and Lifshitz \cite{land-lif}) is to lump $m_0$ and $\eps^{-1} m_e$
together and to take the limits $\eps\to 0$ and $m_0 \to - \infty$
at constant $m_0+\eps^{-1} m_e=m_{exp}$, the experimentally observed
mass of the particle. Then (\ref{portu}) reads as
\begin{equation}\label{readsas}
   m_{exp} \ddot{r} = -\nabla V(r) + a \mbox{\textsf{e}}^{2}
   \stackrel{...}{r}.
\end{equation}
Since this equation is of third order, one needs besides $q^{0},v^{0}$
also ${\dot u}(0)$ as initial condition which has to be extracted somehow
from the initial data of the full system.
Even worse, (\ref{readsas}) has solutions which are
exponentially unbounded in time, the famous run-away
solutions. Thus one needs an additional criterion to single out
the solutions of physical relevance. Dirac \cite{dirac}, and later
Haag \cite{haag}, argued that physical solutions have to satisfy
the asymptotic condition
\begin{equation}\label{as}
   \lim_{t\to\infty} \ddot{r}(t)=0,
\end{equation}
as a substitute for the missing initial condition $\ddot{r}(0)$.
The validity of the asymptotic condition
has been checked only in trivial cases; see \cite{rohr}.
For general $V$ one should expect the solutions to (\ref{portu})
to be chaotic. Physical and unphysical solutions might be badly mixed up.
On a more practical level, the physical solutions are unstable and
therefore difficult to compute numerically. To put it in the words of
W.~Thirring \cite{thirr2}: ``{\it ...(\ref{readsas}) has not only
crazy solutions and there are attempts to separate sense from nonsense
through special initial conditions. But one hopes that the true solution
to the problem will look differently and that the nature of the equations
of motion is not so highly unstable that the act of balance can be
achieved only through a stroke of good fortune in the initial
conditions.}''

This is indeed the case, as we are going to show in this paper,
and our resolution requires just a little twist. If according 
to (\ref{111})
we adopt the macroscopic time scale, then (\ref{portu}) reads
\begin{equation}\label{schallu}
   (m_0 + m_e)\ddot{r} = -\nabla V(r) + \eps a \mbox{\textsf{e}}^{2}
   \stackrel{...}{r}
\end{equation}
which just reflects that radiation reaction is a small correction
to the Hamiltonian motion. The bare mass $m_0$ should be kept
strictly positive. Otherwise, ${\cal H}_{mac}$ from (\ref{Hmac})
is not bounded from below and  (\ref{111}) has solutions increasing
exponentially in time, a phenomenon completely unrelated
to run-away solutions, however.

In (\ref{schallu}) the highest derivative appears with a small prefactor.
Such differential equations are studied in geometric singular
perturbation theory. From there we know that (\ref{schallu}) has a
six-dimensional invariant center manifold ${\cal I}_\eps$,
which is only ${\cal O}(\eps)$ away from the Hamiltonian manifold
${\cal I}_0 = \{(q,\dot{q}, \ddot{q}): (m_0+ m_e) \ddot{q} =
-\nabla V(q)\}$. For initial conditions slightly off ${\cal I}_\eps$
the solution moves away from ${\cal I}_\eps$ exponentially fast.
On ${\cal I}_\eps$, $\dot{q}$ is bounded away from $1$, $\ddot{q}$ is 
bounded,
and the motion is governed by an effective second order equation,
cf.~Eq.~(\ref{second}) below, which gives {\em precisely}
the physical solutions. To establish such a result we have to prove
that the solution to (\ref{111}) stays indeed close to ${\cal I}_\eps$.

In our paper we carry out this program, essentially under the same
conditions as in \cite{KKS}, namely a sufficiently differentiable $V$ and
${|\rho|}_{L^2}$ small. Our main additional estimate is
\begin{equation}\label{ury}
   |\stackrel{...}{v}^{\,\eps} (t)| \le C
\end{equation}
uniformly in $\eps$ and $t\in\R$. Thereby we can bound
one further order in the rigorous Taylor expansion and obtain,
setting $\dot{q}^\eps= v^\eps$,
\begin{equation}\label{ab-form-full}
   m(v^\eps) \dot{v}^\eps = -\nabla V(q^\eps)
   + \eps a(v^\eps)\ddot{v}^\eps + \eps b(v^\eps, \dot{v}^\eps)
   + \eps^2 f^\varepsilon (t),\quad t\ge \eps t_1,
\end{equation}
with $|f^\eps(t)|\le C$ and coefficient functions $a,b$
that will be defined below. Clearly (\ref{ab-form-full}) should be
compared with
\begin{equation}\label{ab-form}
   \dot{r} = u\,,\quad m(u)\dot{u} = -\nabla V(r) + \eps a(u)\ddot{u}
   + \eps b(u, \dot{u}).
\end{equation}
Our crucial observation is that the condition $|u(t)|\le {\rm const}.<1$
for all $t$ holds only on the center manifold
${\cal I}_\eps$. Thus the a priori estimate $|\dot{q}^\eps (t)|
\le\bar{v}<1$,
see (\ref{v-bounds}), together with the initial conditions
$r(0)=q^0$, $u(0)=v^0$, {\it uniquely} singles out that solution
of (\ref{ab-form}) which is to be compared with the true solution.

Since on the error term $f^{\varepsilon}(t)$ in (\ref{ab-form-full})
we only know that it is uniformly bounded, the difference $|q^{\eps}(t)
-r(t)|$, with $r(t)$ having initial conditions on ${\cal I}_\eps$, can be
bounded at best as $\eps e^{ct}$. Thus on the time scale $t={\cal O}(1)$
we seem to be back to the result (\ref{li}) already proved in \cite{KKS}.
To distinguish, from this point of view, between (\ref{ab-form}) and
(\ref{vel}) we would have to control the difference with a precision of
order $\eps^{2}$. At present we do not know whether this is
possible, but nevertheless we can prove the weaker statement
\begin{equation}\label{est}
   |H(q^{\eps}(t),v^{\eps}(t)) - H(r(t),u(t))| \le {\rm const.}\,\eps^{2},
\end{equation}
where $H$ is the energy from (\ref{ener}). Thus on a surface of constant
energy the difference $|q^{\eps}(t) - r(t)|$ could be of order $\eps$,
whereas along $\nabla H$ it must be of order $\eps^{2}$. In
addition to (\ref{est}) it may also be shown that in fact
$|q^{\eps}(t)-r(t)|\sim\eps^3$ on the short time scale
$t={\cal O}(\eps)$, a result that is quite natural from
the viewpoint of singularly perturbed ODEs. On the original
time scale of (\ref{system}) this amounts at least to an estimate with
precision $\eps^2$ over time intervals of length ${\cal O}(1)$,
a result that could not have been obtained from the bounds in \cite{KKS}.


\section{Main results}\label{main-res}
\setcounter{equation}{0}

We give some more details and state our main results precisely.
First we have to establish the bound (\ref{ury}).
\begin{lemma}\label{q4-bound} For ${|\rho|}_{L^2}$ sufficiently small
we have
\[ \sup_{t\in\R}|\stackrel{...}{v}^{\eps}(t)|\le C \]
for every solution of (\ref{111}) which starts on the soliton manifold
${\cal S}$. Both the constant $C$ and the bound for ${|\rho|}_{L^2}$
depend only on the initial data.
\end{lemma}
The bound of Lemma 2.1 may be used to Taylor expand the self-force
\begin{equation}\label{Fs-def}
   F_s^{\eps}(t)=\sqrt{\eps}\,\int d^3x\,\phi(x, t)\nabla
   \rho_{\eps}(x-q^{\eps}(t))
\end{equation}
in (\ref{111}) as
\begin{equation}\label{self-form}
   F_s^{\eps}(t)=-m_f(v^{\eps}(t))\dot{v}^{\eps}(t)
   + \eps a(v^\eps(t))\ddot{v}^\eps(t) + \eps b(v^\eps(t),
   \dot{v}^\eps(t)) + {\cal O}(\eps^2)\,,\quad t\ge \eps t_1\,,
\end{equation}
which together with the second equation in (\ref{111}) yields 
(\ref{ab-form-full}).
Here $m_f$ is defined in (\ref{mf-def}),
and $t_1= 2R_\rho/(1-\bar{v})$ is the microscopic time
the wave equation needs to forget its data because of the
compact support of $\rho$ and the velocity bound, cf.~assumption $(C)$
and (\ref{v-bounds}). The coefficient functions are given by
\begin{eqnarray}
   a(v)\ddot{v} & = & (\e^2/24\pi)
   (\ddot{v}\cdot\nabla_v)\nabla_v\gamma^{2} =
   \,(\e^2/12\pi)[\gamma^{4}\ddot{v} +
   4\gamma^{6}(v\cdot\ddot{v})v]\, , \label{a-def} \\
   b(v, \dot{v}) & = & (\e^2/32\pi)
   {(\dot{v}\cdot\nabla_v)}^2\nabla_v \gamma^{2}
   \nonumber\\ & = & (\e^2/4\pi)[2\gamma^{6}(v\cdot\dot{v})\dot{v}
   + \gamma^{6}\dot{v}^2 v + 6\gamma^{8}(v\cdot\dot{v})^2 v],
   \label{b-def}
\end{eqnarray}
$\dot{v},\ddot{v}\in\R^3$, with $\gamma = 1/\sqrt{1-v^{2}}$, $|v|<1$.

Next we explain the existence and the role of the center-like
manifolds ${\cal I}_\eps$ in greater detail. We refer to
\cite{saka, jones} for further background on geometric singular
perturbation theory. To rewrite (\ref{ab-form}) as
a singular perturbation problem, let
\begin{eqnarray*}
   & x=(r, u)\in\R^3\times {\cal V}\,,\quad y=\dot{u}\in\R^3\,,\quad
   f(x, y)=(x_2, y)\in {\cal V}\times\R^3\,,\quad\mbox{and} & \\[0.5ex]
   & g(x, y, \eps)= {a(x_2)}^{-1}[m(x_2)y+\nabla V(x_1)
   -\eps b(x_2, y)]\,. &
\end{eqnarray*}
Then (\ref{ab-form}) reads as
\begin{equation}\label{singul}
   \dot{x}=f(x, y)\, ,\quad \eps\dot{y}=g(x, y, \eps)\, .
\end{equation}
We intend to apply the results from \cite{saka} to (\ref{singul})
in order to find a center-like manifold for the perturbed problem near the
corresponding manifold for the $(\eps=0)$-problem. With
$h(x)=-m(x_2)^{-1}\nabla V(x_1)$, let
\begin{eqnarray}\label{C-def}
   {\cal I}_0 & = & \{(x, y): g(x, y, 0)=0\}
   = \{(r, u, \dot{u}): m(u)\dot{u}=-\nabla V(r)\} \nonumber\\
   & = & \{(x, h(x)): x\in\R^3\times {\cal V}\}
\end{eqnarray}
be this invariant manifold for (\ref{singul}) with $\eps=0$. The
flow on ${\cal I}_0$ is governed by the equation $\dot{x}=f(x, h(x))$,
or stated differently, $m(\dot{r})\ddot{r}=-\nabla V(r)$,
the familiar Hamiltonian flow.

To see that ${\cal I}_0$ is perturbed to some ${\cal I}_\eps$
with $\eps$ small, we have to modify
the functions $a(u)$, $m(u)$, and $b(u, \dot{u})$ for
$|u|$ close to one due to the singularity at $|u|=1$.
This will cause no problems later on, since we already have the
a priori bound $|v^{\eps}(t)|\le\bar{v}<1$ for the velocity of the true
system. In (\ref{deltabar-def}) below, we will fix a small
$\bar{\delta}=\bar{\delta}(\bar{v})>0$ satisfying some estimates;
$\bar{\delta}$ depends only on bounds for the initial data, since
$\bar{v}$ does so. Let
\[ {\cal K}_{1-\bar{\delta}}
   =\R^3\times\{u\in\R^3: |u|\le 1-\bar{\delta}\}, \]
We continue $a(u)$, $m(u)$, and
$b(u, \dot{u})$ with their values at $|u|=1-\bar{\delta}$
to the missing infinite strip $1-\bar{\delta}<|u|<1$.
Then the basic assumptions $(I)$, $(II)$ from \cite[p.~45]{saka}
are satisfied, since ${\cal I}_0$ is also what is called normally
hyperbolic, i.e.~repulsive in the direction normal
to ${\cal I}_0$ at an $\eps$-independent rate, see Lemma \ref{hyp} below.
Hence we find $\eps_0=\eps_0(\bar{\delta})>0$ and a $C^1$-function
$h(x, \eps)=h_\eps(x): \R^3\times {\cal V}\times ]0, \eps_0]\to\R^3$
such that for $\eps\le\eps_0$,
\[ {\cal I}_\eps=\{(x, h_\eps(x)): x\in\R^3\times {\cal V}\} \]
is forward invariant for the flow  (\ref{ab-form}) with the
modified functions $a, m,b$. Since the modified equation
agrees with (\ref{ab-form}) in the interior of ${\cal K}_{1-\bar{\delta}}$,
we conclude that ${\cal I}_\eps$ is locally invariant for the flow
(\ref{ab-form}), i.e. the solution of the modified equation
is the solution to the original equation as long as it
does not reach the boundary set
$\{(x, h_\eps(x))=(r, u, h_\eps(r, u)): |u|=1-\bar{\delta}\}$.
The flow for $\eps=0$ is then perturbed  to
$\dot{x}=f(x, h_\eps(x))$ for $\eps\le\eps_0$.

We will show in Theorem \ref{ZM-thm-U} below that for
$\eps\in ]0, \eps_1]$, with $\eps_1>0$ sufficiently small,
all solutions of (\ref{ab-form}) starting at points
$(r, u, h_\eps(r, u))\in {\cal I}_\eps$ with
$|u|\le\bar{v}$, will indeed stay away from the boundary
$\{(r, u, h_\eps(r, u)): |u|=1-\bar{\delta}\}$ for all future times.
In addition, $\nabla V(r(t))\to 0$ and $\ddot{r}(t)\to 0$ as $t\to\infty$,
which is just the asymptotic condition (\ref{as}) postulated by
Dirac and Haag. If the potential is sufficiently confining,
then the solution trajectory on ${\cal I}_\eps$ not only approaches
the set of critical points for $V$ in the long-time limit,
but it converges to some definite critical point. Moreover,
we will show that for all solutions on the center manifold,
$\dot{u}(t)$ and $\ddot{u}(t)$ are bounded, and $u(t)$ is bounded 
away from $1$,
uniformly in $\eps$ and $t$. Conversely, every such solution to
(\ref{ab-form}) has to lie on ${\cal I}_\eps$.
Thus ${\cal I}_\eps$ indeed characterizes the physical solutions.

To summarize, we have established now the existence of a center
manifold ${\cal I}_\eps$ with a well-defined (semi-) flow
on it that gives a unique solution to (\ref{ab-form}) for
initial velocities bounded by $\bar{v}$.

For the potential $V\in C^3(\R^3)$ we assume that it is bounded in 
the sense
$\inf_{q\in\R^3} V(q)>-\infty$ and
$$
\sup_{q\in\R^3}\,\Big(|V(q)|+|\nabla V(q)|+|\nabla \nabla V(q)|
+|\nabla\nabla\nabla V(q)|\Big)
<\infty \,.\eqno{(U)}
$$
The method works equally well for $V\in C^3(\R^3)$ which is
confining, i.e.,
$$
V(q)\to\infty \quad{\rm as}\quad |q|\to\infty\,,
\eqno{(U')}
$$
as will be made more precise in Section \ref{center-sect},
cf.~Theorem \ref{ZM-thm-Us}.

Our main result is the following

\begin{theorem}\label{mainthm}
Assume $(U)$ or $(U')$ for the potential, and let the initial data
$(\phi^0(x), \pi^0(x), q^0, v^0)$ for (\ref{111}) be given by (\ref
{macro-ini}). Let ${|\rho|}_{L^2}$ and $\eps\le\eps_1$ be 
sufficiently small,
and introduce the center manifolds ${\cal I}_\eps$
for the comparison dynamics (\ref{ab-form}) as explained above. At time
$\eps t_1 = \eps 2R_\rho/(1-\bar{v})$ we match
the initial values, $r(\eps t_1)=q^{\eps}(\eps t_1)$, $u(\eps t_1)
=v^\eps(\eps t_1)$, for the motion on the center manifold, i.e.,
the initial data for the comparison dynamics are
\[ (q^\eps(\eps t_1), v^\eps(\eps t_1), h_\eps(q^\eps(\eps t_1),
v^\eps(\eps t_1)))
   \in {\cal I}_\eps\,. \]
Then for every $\tau>0$ there exists $c(\tau)>0$ such that for all
$t\in [\eps t_1, \eps t_1+\tau]$
\begin{equation}\label{calv}
   |q^\eps(t)-r(t)|\le c(\tau)\eps\,,\quad
   |v^\eps(t)-u(t)|\le c(\tau)\eps\,,\quad\mbox{and}\quad
   |\dot{v}^\eps(t)-\dot{u}(t)|\le c(\tau)\eps.
\end{equation}
In addition we have the bound
\begin{equation}\label{Hbound}
   |H(q^{\eps}(t),v^{\eps}(t)) - H(r(t),u(t))| \le c(\tau)\eps^{2}.
\end{equation}
\end{theorem}
\smallskip

\begin{remarks}\label{AW}{\rm
(i) As already mentioned at the end of the introduction, we can also show
\begin{equation}\label{lis-pk}
   |q^\eps(t)-r(t)|\le c(\tau)\eps^3\quad\mbox{and}\quad
   |v^\eps(t)-u(t)|\le c(\tau)\eps^2
\end{equation}
for $t\in [\eps t_1, \eps t_1+\eps\tau]$, i.e., $t={\cal O}(\eps)$,
cf. Proposition \ref{besser-bd}.
\medskip

\noindent
(ii) The construction of the center manifolds
and the upper bound for ${|\rho|}_{L^2}$ rely only on bounds
for the data, but not on properties of a particularly chosen
solution. Our main technical assumption is a sufficiently
small ${|\rho|}_{L^2}$ which is presumably not necessary.
\medskip

\noindent
(iii) In \cite{KKS} we did not require the true
solution to start on the soliton manifold, but instead
to start close to it. We refer the criterion \cite[Thm.~2.6]{KKS}
for an ``adiabatic'' family of solutions. The same generality
could be achieved in the present context, using an
appropriately modified version of \cite[Thm.~2.6]{KKS}.
In Section \ref{append} we derive the relevant estimates,
in particular (\ref{data-null}), in full generality
containing a non-zero initial difference $Z(0)$.
The corresponding generalization of Theorem \ref{mainthm} is then
straightforward.
However, since we did not want to obscure our main achievement
through technicalities, we decided to elaborate here
the more accessible case of a trajectory starting right on the
soliton manifold. In the same spirit we do not consider
arbitrary time intervals of length $\tau$, but only
the particular $[\eps t_1, \eps t_1+\tau]$. \medskip

\noindent
(iv) The existence of solutions to (\ref{system}) is
discussed in \cite[Lemma 2.2]{KKS}. For every initial value
$Y^0=(\phi^0(x), \pi^0(x), q^0, p^0)\in {\cal E}$ we find a unique
(weak) solution $Y(\cdot)\in C(\R, {\cal E})$ such that
$Y(0)=Y^0$. Here the state space is
${\cal E}=D^{1, 2}(\R^3)\oplus L^2(\R^3)\oplus\R^3\oplus\R^3$
[where $D^{1, 2}(\R^3)=\{\phi\in L^6(\R^3): |\nabla\phi|\in
L^2(\R^3)\}$] with norm $|Y|_{{\cal E}}={|\nabla\phi|}_{L^2}
+{|\pi|}_{L^2}+|q|+|p|$.}
\end{remarks}

Having such fairly precise information on the particle
trajectory we can also determine the adiabatic limit $\eps\to 0$
of the fields $(\phi, \pi)$ in (\ref{111}) through
the solution of the inhomogeneous wave equation.
We generate the initial data as explained in the
introduction. On the level of the comparison dynamics this
means to extend $r(t)$ and $u(t)$ to negative times $t\le 0$ by
$r(t)= q^0+t v^0$ resp.~$u(t)=v^0$. Let the
retarded time $t_{{\rm ret}}$, depending on $x$ and $t$, be the 
unique solution
of $t_{{\rm ret}} = t-|x-r(t_{{\rm ret}})|$, and let $\hat{n}(x,t) =
(x-r(t_{{\rm ret}}))/|x-r(t_{{\rm ret}})|$.
\begin{theorem}\label{feld-thm}
Under the conditions of Theorem \ref{mainthm} and for
the fields $(\phi, \pi)$ from (\ref{111}) we have
for $x\neq r(t)$ the pointwise limits
\begin{equation}\label{phi-lim}
   \lim_{\eps \to 0} \,\frac{1}{\sqrt{\eps}}\,\phi(x,t)
   = -\frac{\e}{4\pi|x-r(t_{{\rm ret}})|}\,{\Big(1-\hat{n}(x,t) \cdot
   u(t_{{\rm ret}})\Big)}^{-1}
\end{equation}
and, except for the light cone $\{x: |x|=t>0\}$,
\begin{eqnarray}
   \lefteqn{\lim_{\eps \to 0}\,\frac{1}{\sqrt{\eps}}\,
   \pi (x,t)}\nonumber\\ & = & - \frac{\e}{4\pi|x-r(t_{{\rm ret}})|} \,
   {\Big(1-\hat {n}(x,t) \cdot u(t_{{\rm ret}})\Big)}^{-3}\,
   \hat{n}(x,t)\cdot \dot{u}(t_{{\rm ret}}) \nonumber \\
   & & - \frac{\e}{4\pi|x-r(t_{{\rm ret}})|^2} \,
   {\Big(1-\hat{n}(x,t) \cdot u(t_{{\rm ret}})\Big)}^{-3}\,
   (\hat{n}(x,t)\cdot u(t_{{\rm ret}}) - {u(t_{{\rm ret}})}^2).
   \nonumber\\ & & \label{pi-lim}
\end{eqnarray}
\end{theorem}

The paper is organized as follows. Since the proof of Lemma
\ref{q4-bound} is rather technical, we moved it to an appendix,
Section \ref{append}. The derivation of the representation
(\ref{self-form}) of the self-force term is the contents of
Section \ref{self-sect}. In Section \ref{center-sect} we
give supplementary remarks on the behaviour of solutions on the
center manifold, whereas in Section \ref{main-sect} we carry out
the proofs of Theorem \ref{mainthm} and Proposition
\ref{besser-bd}. Section \ref{feld-sect} contains the
proof of Theorem \ref{feld-thm}, and finally in Section
\ref{dissi-sect} we determine the amount of energy radiated
to infinity.


\section{Representation of the self-force}\label{self-sect}
\setcounter{equation}{0}

In this section we show that the self-force $F_s^\eps(t)$ from 
(\ref{Fs-def})
can be written in the form (\ref{self-form}). We carry out this
computation on the original
fast time scale corresponding to (\ref{system}) since we will need
some of the arguments from \cite{KKS}. Thus we consider
\[ F_s(t)=\int d^3x\,\phi(x, t)\nabla\rho (x-q(t)). \]
Since $\phi(x, t)=\phi_r(x, t)+\phi_0(x, t)$, where
$\ddot{\phi}_0=\Delta\phi_0$ with the initial values $\phi_0(x, 0)=
\phi^0(x)$
and $\pi_0(x, 0)=\pi^0(x)$, and since
\[ \phi_r(x, t)=-\frac{1}{4\pi}\int_0^t\,\frac{ds}{t-s}
   \,\int_{|y-x|=t-s} d^2y\,\rho(y-q(s)) \]
is the retarded potential, we can decompose accordingly,
\[ F_s(t)=F_0(t)+F_r(t)=\langle\phi_0(\cdot, t),
   \nabla\rho(\cdot-q(t))\rangle
   + \langle\phi_r(\cdot, t), \nabla\rho(\cdot-q(t))\rangle\,. \]

\begin{lemma}\label{regh} The function $F_0(t)$ vanishes for
$t\ge t_1=\displaystyle 2R_\rho/(1-\bar{v})$.
\end{lemma}
{\bf Proof\,:} Let $U(t)$ denote the group generated by the free
wave equation in $D^{1, 2}(\R^3)\oplus L^2(\R^3)$. Then
(\ref{ini-cond}) and Fourier transformation implies
\[ (\phi^0(x), \pi^0(x))
   =-\int_{-\infty}^0 ds\,[U(-s)\bar{\rho}(\cdot-q^0-v^0s)](x) \]
with $\bar{\rho}(x)=(0, \rho(x))$. Thus Kirchhoff's
formula yields, as a consequence of $|v^0|<1$, that
$\phi_0(x, t)=0$ for $|x-q^0|\le t-R_\rho$. Since
$|q(t)-q^0|\le \bar{v}t$, the claim follows.
{\hfill $\Box$}\bigskip

Hence to show (\ref{self-form}) it is enough to prove

\begin{lemma} For $t\ge t_1$,
\[ F_r(t)=-m_f(v(t))\dot{v}(t)
   + a(v(t))\ddot{v}(t) + b(v(t), \dot{v}(t)) + {\cal O}(\eps^3)\, , \]
cf.~(\ref{mf-def}), (\ref{a-def}), and (\ref{b-def}).
\end{lemma}
{\bf Proof\,:} We follow the proof of \cite[Lemma 5.1]{KKS}, but
expand
\[ q(s)=q(t)-v(t)(t-s)+\frac{1}{2}\dot{v}(t)(t-s)^2-\frac{1}{6}
\ddot{v}(t)
   (t-s)^3+{\cal O}(\eps^3) \]
up to third order, which is allowed by Lemma \ref{q4-bound}. Through
Fourier transformation we arrive at
\begin{eqnarray*}
   F_r(t) & = & (-i)\int_0^t\,ds\int d^3k\,|\hat{\rho}(k)|^2 
   \frac{k}{|k|}\,
   \sin |k|(t-s)\,e^{-i (k\cdot v) (t-s)}\,\\ & & \hspace{11em}\times
   e^{-i [-\frac{1}{2}(k\cdot\dot{v})(t-s)^2+\frac{1}{6}(k\cdot\ddot{v})
   (t-s)^3]}+{\cal O}(\eps^3)\,,
\end{eqnarray*}   
with $v=v(t)$, etc.. As in \cite[Lemma 5.1]{KKS},
here and in the following $\int_0^t ds (\ldots)$ can be changed forth and
back to $\int_{t-T}^t ds (\ldots)$ for all $t, T\ge t_1$. Because
\begin{eqnarray*}
  e^{-i [-\frac{1}{2}(k\cdot\dot{v})(t-s)^2+\frac{1}{6}(k\cdot\ddot{v})
   (t-s)^3]} & = & 1+\frac{i}{2}(k\cdot\dot{v})(t-s)^2-\frac{i}{6}(k\cdot
   \ddot{v})(t-s)^3 \\ & & -\frac{1}{8}(k\cdot\dot{v})^2 (t-s)^4+
   {\cal O}(\eps^3)
\end{eqnarray*}   
for $t-s={\cal O}(1)$ by (\ref{q23-bounds}) below,
we obtain, for $t, T\ge t_1$,
\begin{eqnarray*}
   F_r(t) & = & (-i)\int_0^T\,d\tau\int d^3k\,|\hat{\rho}(k)|^2 
   \frac{k}{|k|}\,
   \sin |k|\tau\,e^{-i (k\cdot v)\tau}\,
   \\ & & \hspace{4em}\times \Big[1+\frac{i}{2}(k\cdot\dot{v})\tau^2
   -\frac{i}{6}(k\cdot\ddot{v})\tau^3
   -\frac{1}{8}(k\cdot\dot{v})^2 \tau^4\Big] + {\cal O}(\eps^3)\,.
\end{eqnarray*}   
Let
\[ I_p=\int_0^T\,d\tau\,\frac{\sin |k|\tau}{|k|}\,
   e^{-i (k\cdot v)\tau}\tau^p\,,\quad p=0, \ldots, 4\,. \]
Then
\[ (\ddot{v}\cdot\nabla_v)\nabla_v I_1=-k(k\cdot\ddot{v})I_3
   \quad\mbox{and}\quad
   (\dot{v}\cdot\nabla_v)^2\nabla_v I_1=ik(k\cdot\dot{v})^2 I_4\, . \]
Our claim now follows from Lemma \ref{I1} below,
$\int d^3k\,|\hat{\rho}(k)|^2 k I_0\to 0$, and
$(1/2)\int d^3k\,|\hat{\rho}(k)|^2 k (k\cdot\dot{v})I_2
\to -m_f(v(t))\dot{v}(t)$ for $T\to\infty$; see \cite[Appendix A]{KKS}.
{\hfill $\Box$}\smallskip

\begin{lemma}\label{I1} We have the identity
\[ \int_0^\infty dt\,t \int d^3k |\hat{\rho}(k)|^2
   \frac{\sin |k| t}{|k|}\,e^{-i (k\cdot v) t}
   = (\mbox{{\sf e}}^{2}/4\pi)\gamma^{2}\, . \]
\end{lemma}
{\bf Proof\,:} Since $\hat{\rho}(k)=\hat{\rho}_r(|k|)$
is radial, and by transformation to polar coordinates,
\[ \int d^3k |\hat{\rho}(k)|^2
   \frac{\sin |k|t}{|k|}\,e^{-i (k\cdot v) t}
   = \frac{4\pi}{t|v|}\int_0^\infty dR\,{|\hat{\rho}_r(R)|}^2
   \sin(Rt)\sin(Rt|v|)\,. \]
Thus for fixed $T>0$,
\begin{eqnarray*}
  \lefteqn{\int_0^T dt\,t \int d^3k |\hat{\rho}(k)|^2
   \frac{\sin |k| t}{|k|}\,e^{-i (k\cdot v) t}} \\
   & = & \frac{2\pi}{|v|}\int_0^\infty \frac{dR}{R}
   {|\hat{\rho}_r(R)|}^2\,\bigg( \frac{\sin(R(1-|v|)T)}{1-|v|}
   -\frac{\sin(R(1+|v|)T)}{1+|v|}\bigg)\,.
\end{eqnarray*}   
To complete the proof we only need to verify that
$\int d^3k |\hat{\rho}(k)|^2\,|k|^{-3}\sin |k|T\to \mbox{{\sf e}}^{2}/4\pi$
as $T\to\infty$. To see this, let $\hat{\psi}(k)=|k|^{-3}\sin (|k|T)$.
Then
\[ \int d^3k |\hat{\rho}(k)|^2\,\hat{\psi}(k)={(2\pi)}^{-3/2}
   \int d^3x\rho(x)\int d^3y\,\rho(y)\psi(x-y) \]
and we are going to show $\psi(x)\to\sqrt{\pi/2}$ as $T\to\infty$. We have,
by transformation to polar coordinates,
\[ (2\pi)^{3/2}\psi(x) = \int d^3k\,\hat{\psi}(k)e^{-i k\cdot x}
   = 4\pi\,\int_0^\infty ds\,\frac{\sin(s)}{s}\,
   \frac{\sin(s|x|/T)}{s|x|/T}\to 2\pi^2 \]
for $T\to\infty$. This completes the proof. {\hfill$\Box$}\bigskip


\section{More about the center manifold}\label{center-sect}
\setcounter{equation}{0}

In this section we explain the behaviour
of solutions on the center manifold. First we show that the
unperturbed manifold ${\cal I}_0$ from (\ref{C-def})
is hyperbolic in normal direction.

\begin{lemma}\label{hyp} The eigenvalues of $D_y g(x, y, 0)=
	a(x_2)^{-1}m(x_2)$
are bounded below by a positive constant, uniformly in $x=(r, u)$ with
$r\in\R^3$ and $|u|\le 1-\delta$, for all prescribed $\delta\in ]0, 1]$.
\end{lemma}
{\bf Proof\,:} By \cite[Thm.~2, p.~185]{lanc}, $a(u)$ and $m(u)$ can be
simultaneously transformed to diagonal form through a single
non-singular matrix $B$. In addition,  denoting by $b_j\neq 0$
the $j$th column of $B$ and by $\lambda_j$  the $j$th
eigenvalue of $a(u)^{-1}m(u)$,  one has
$\lambda_j a(u)b_j=m(u)b_j$, $j=1, 2, 3$.
Multiplication by $b_j$ leads to $\lambda_j
(\e^2/12\pi)\gamma^3 [\gamma b_j^2+4\gamma^3 {(v\cdot b_j)}^2]
\ge\gamma b_j^2+\gamma^3 {(v\cdot b_j)}^2$, and thus
$\lambda_j\ge (3\pi/\e^2)\gamma^{-3}$. {\hfill$\Box$}\bigskip

Since $a(u)$, $m(u)$ are modified to be constant outside
$|u|\le 1-\bar{\delta}$, their corresponding eigenvalues
are uniformly bounded below for $|u|<1$. As a consequence
of Lemma \ref{hyp} the manifolds ${\cal I}_{\eps}$
are unstable at some exponential rate $e^{\mu t}$
for solutions in the normal direction.

We note that, by \cite[Thm.~2.1]{saka},
\begin{equation}\label{h-schranke}
   \sup\{|h_\eps(r, u)|: (r, u)\in\R^3\times {\cal V},
   \eps\in ]0, \eps_0]\}\le c=c(\bar{\delta}).
\end{equation}
Our next aim is to prove global existence
of solutions to (\ref{ab-form}) forward in time which start over
${\cal K}_{\bar{v}}=\R^3\times\{u\in\R^3: |u|\le\bar{v}\}$ on the
center manifold, provided $\eps\le\eps_1$ with
$\eps_1>0$ sufficiently small. For this purpose
we introduce a suitable Lyapunov function.

\begin{lemma}\label{G-def}
Let
\[ G_\eps(r, u, \dot{u})=H(r, u)
   -\eps (a(u)\dot{u})\cdot u = {\cal E}_s(u)+V(r)-\eps (a(u)\dot{u})
   \cdot u\,. \]
Then along solutions $(r(t), u(t), \dot{u}(t))$ of (\ref{ab-form}) we have
\begin{equation}\label{lya}
   \frac{d}{dt} G_\eps(r, u, \dot{u})=-\eps (\e^2/12\pi)\,
   [6\gamma^8 (u\cdot\dot{u})^2 + \gamma^6\dot{u}^2].
\end{equation}
\end{lemma}
{\bf Proof\,:} Observing that
\[ (a(u)\dot{u})\cdot u=(\e^2/12\pi)\gamma^6\,(1+3u^2)
(u\cdot\dot{u}), \]
this is a straightforward calculation. {\hfill$\Box$}\bigskip

Through the Lyapunov function $G_{\eps}$ we can control
the long time behaviour.

\begin{theorem}\label{rigo} Let $(U)$ or $(U')$ hold and
let any global solution $(r(t), u(t))$ of (\ref{ab-form}) be given 
such that
$\sup_{t\ge 0}|u(t)|\le\bar{u}(\eps)<1$ and
$\sup_{t\ge 0}|\dot{u}(t)|\le c(\eps)$, for possibly
$\eps$-dependent constants $\bar{u}(\eps)$ and $c(\eps)$.
Then
\[ \dot{u}(t)\to 0,\quad\ddot{u}(t)\to 0,\quad\mbox{and}\quad
   \nabla V(r(t))\to 0\quad\mbox{as}\quad t\to\infty. \]
\end{theorem}
{\bf Proof\,:} Denoting by $c(\eps)$ or $C(\eps)$ general $\eps$-dependent
constants, by Lemma \ref{G-def} we have  along a trajectory
\begin{eqnarray*}
   c(\eps)\,\int_0^T \dot{u}^2\,dt & \le &
   - \int_0^T\,\frac{d}{dt}G_\eps\,dt
   \\ & = & - {\cal E}_s(u(T))-V(r(T))+\eps (a(u(T))\dot{u}(T))\cdot u(T)
   \\ & & + {\cal E}_s(u^0)+V(r^0)-\eps (a(u^0)\dot{u}^0)\cdot u^0
   \\ & \le & C(\eps, \mbox{data}).
\end{eqnarray*}
For the last estimate observe $\inf_{r\in\R^3} V(r)>-\infty$ in both
cases $(U)$ and $(U')$. Thus $\int_0^\infty\dot{u}^2\,dt\le
C(\eps, \mbox{data})$ and, by (\ref{ab-form}), also
$\sup_{t\ge 0}|\ddot{u}(t)|\le C(\eps, \mbox{data})$.
Hence we conclude $\dot{u}(t)\to 0$ as $t\to\infty$. Next,
differentiation of (\ref{ab-form}) yields
$\sup_{t\ge 0}|\stackrel{...}{u}(t)|\le
C(\eps, \mbox{data})$, and thus from $\dot{u}(t)\to 0$ we find
$\ddot{u}(t)\to 0$. Therefore $\nabla V(r(t))\to 0$ follows from
the equation (\ref{ab-form}). {\hfill$\Box$}\bigskip

In the demonstration of the following theorem we  use the
sublevel sets $\{G_\eps\le c\}=\{(r, u, \dot{u}): G_\eps(r, u, \dot{u})
\le c\}$ and $\{H\le c\}=\{(r, u): H(r, u)\le c\}$ for
$c\in\R$. However, before proceeding, we first have to introduce
an appropriate $\bar{\delta}=\bar{\delta}(\bar{v})>0$ small to
modify the functions $a(u)$, $m(u)$, and $b(u, \dot{u})$
outside $|u|\le 1-\bar{\delta}$, cf.~Section \ref{main-res}. To do this,
we assume $(U)$ from now on. The case $(U')$ is discussed
in the remarks below.
Since $V$ is bounded and $\bar{v}<1$, we can find $c_0\in\R$ such that
${\cal K}_{\bar{v}}\subset \{H\le c_0\}$. Then as a consequence of
${\cal E}_s(u)\to\infty$ for $|u|\to 1$, we have
\begin{equation}\label{s0-def}
   s_0=\sup\Big\{|u|: (r, u)\in\{H\le c_0+1\}
   \,\,\mbox{for\,\,some\,\,} r\in\R^3\Big\}<1\,.
\end{equation}
Let us define
\begin{equation}\label{deltabar-def}
   \bar{\delta}=\min\{(1-\bar{v})/2, (1-s_0)/2\}>0.
\end{equation}

\begin{theorem}\label{ZM-thm-U}
Assume the potential $V$ to satisfy the condition $(U)$. Then there exists 
$\eps_1>0$
depending only upon $\bar{v}$ such that for $\eps\in ]0, \eps_1]$
all solutions of (\ref{ab-form}) starting at points
$(r, u, h_\eps(r, u))\in {\cal I}_\eps$, $|u|\le\bar{v}$,
stay away from the boundary $\{(r, u, h_\eps(r, u)): 
|u|=1-\bar{\delta}\}$
for all future times. In particular, solutions exist globally.
\end{theorem}
{\bf Proof\,:} Let us denote the bound $c(\bar{\delta})$
from (\ref{h-schranke}) by $c_1$ and let us fix $c_a>0$
such that $|a(u)|\le c_a$ for all $|u|< 1$. We recall that
$a(u)$ was modified to be constant outside $|u|\le 1-\bar{\delta}$.
We define $\eps_1=\min\{\eps_0, (2c_a c_1)^{-1}\}>0$.

Let $(r, u)\in {\cal K}_{\bar{v}}$. Then $G_\eps(r, u, h_\eps(r, u))
=H(r, u)-\eps (a(u)h_\eps(r, u))\cdot u\le c_0+c_a c_1\eps$.
Because of Lemma \ref{G-def} the set $\{G_\eps\le c_0+c_a c_1\eps\}$
is forward invariant and the solution remains in this set
for all future times. On the other hand, since
$\bar{v}\le 1-2\bar{\delta}<1-\bar{\delta}$,
the solution of the modified problem is a solution
to (\ref{ab-form}) and stays on ${\cal I}_\eps$,
at least for a short times. For the fixed time span where this holds
the solution is of the form $(r_1, u_1, h_\eps(r_1, u_1))$
and we have $H(r_1, u_1)=G_\eps(r_1, u_1, h_\eps(r_1, u_1))
+\eps (a(u_1)h_\eps(r_1, u_1))\cdot u_1\le c_0+c_a c_1\eps+c_a c_1\eps
= c_0+2c_a c_1\eps\le c_0+1$ for $\eps\le\eps_1$.
Therefore by (\ref{s0-def}), $|u_1|\le s_0\le 1-2\bar{\delta}
<1-\bar{\delta}$. This argument shows that in fact the solution
is confined to $\{(r, u, \dot{u}): |u|\le 1-2\bar{\delta}\}$.
Hence the solution of the modified problem exists, is a solution
to (\ref{ab-form}), and stays on ${\cal I}_\eps$
for all future times. {\hfill$\Box$}\bigskip

\begin{cor}\label{Q-bounds} In the setting of Theorem \ref{ZM-thm-U},
for solutions of (\ref{ab-form}) starting on ${\cal I}_\eps$,
\begin{eqnarray}
   & \sup \{|u(t)|: t\in\R, \eps\in ]0, \eps_1]\}
   \le 1-2\bar{\delta}<1\,,\quad\mbox{and}\quad & \nonumber \\
   & \sup\{|\dot{u}(t)|: t\in\R, \eps\in ]0, \eps_1]\}
   + \sup \{|\ddot{u}(t)|: t\in\R, \eps\in ]0, \eps_1]\}
   \le c(\bar{\delta}). & \nonumber\\ & & \label{Q-U-bounds}
\end{eqnarray}
In particular by Theorem \ref{rigo}
\[ \dot{u}(t)\to 0,\quad\ddot{u}(t)\to 0,\quad\mbox{and}\quad
   \nabla V(r(t))\to 0\quad\mbox{as}\quad t\to\infty. \]
\end{cor}
{\bf Proof\,:} The first estimate was mentioned already in the
preceding proof. For the second we note that  (\ref{h-schranke})
applies, since the trajectory stays on the center manifold, $\dot{u}(t)
=h_\eps(r(t), u(t))$. Concerning the last bound, we may write
\begin{equation}\label{1ord-darst}
   h_\eps(r, u)=-{m(u)}^{-1}\nabla V(r)
   +h_{1, \eps}(r, u)\quad\mbox{with}\quad
   |h_{1, \eps}(r, u)|\le c(\bar{\delta})\eps
\end{equation}  
for $(r, u)\in\R^3\times {\cal V}$,
see \cite[Thm.~2.9]{saka}. By (\ref{ab-form}),
\[ |\eps\ddot{u}|\le |{a(u)}^{-1}||m(u)h_{1, \eps}(r, u)
   -\eps b(u, \dot{u})|\le c(\bar{\delta})\eps, \]
so we are done. {\hfill$\Box$}\bigskip

Solutions on ${\cal I}_\eps$ are uniformly bounded,
in the sense of the corollary; in general a bound
on $r(t)$ cannot be expected, e.g.~in a scattering situation.
Conversely, as to be shown next, solutions with uniformly bounded
$u(t)$, $\dot{u}(t)$, and $\ddot{u}(t)$ are confined
to the center manifolds.

\begin{proposition}\label{kiba} Suppose we have a family
$(r^\eps(t), u^\eps(t))$, $\eps\in ]0, \eps_2]$, of solutions
to (\ref{ab-form}) such that
\begin{eqnarray*}
   & \sup \{|u^\eps(t)|: t\in\R, \eps\in ]0, \eps_2]\}
   \le\bar{u}<1\,,\quad\mbox{and}\quad & \\
   & \sup\{|\dot{u}^\eps(t)|: t\in\R, \eps\in ]0, \eps_2]\}
   + \sup \{|\ddot{u}^\eps(t)|: t\in\R, \eps\in ]0, \eps_2]\}
   \le c_2. &
\end{eqnarray*}
Then for sufficiently small $\eps$ the solutions have to lie on
${\cal I}_\eps$.
\end{proposition}
{\bf Proof\,:} Note that we can construct ${\cal I}_\eps$ here
by modifying $a(u), m(u)$, and $b(u, \dot{u})$ to be constant outside,
say, $\{u: |u|\le (1+\bar{u})/2\}$. According to \cite[Thm.~2.1 (ii)]
{saka} there exists $\delta>0$ such that for all $\eps$ small and
solutions $(x(t), y(t))$ to (\ref{singul}) the condition
$\sup_{t\in\R}|y(t)-h(x(t))|\le\delta$ implies that the solution
has to lie on ${\cal I}_\eps$. With $x(t)=(r^\eps(t), u^\eps(t))$ and
$y(t)=\dot{u}^\eps(t)$, this condition is verified since
we obtain from (\ref{ab-form}) and the assumed bounds
$|\dot{u}^\eps(t)+m(u^\eps(t))^{-1}\nabla V(r^\eps(t))|\le
c\eps\le\delta$, the latter for $\eps$ small. {\hfill$\Box$}\bigskip

The asymptotic condition, $\ddot{r}(t)\to 0$,
of Dirac and Haag is also sufficient for a solution
to lie on ${\cal I}_\eps$, in the following sense.

\begin{proposition} Suppose a family $(r^\eps(t), u^\eps(t))$,
$\eps\in ]0, \eps_2]$, of solutions to (\ref{ab-form}) is given
such that
\[ \sup\{|u^\eps(t)|: t\in\R, \eps\in ]0, \eps_2]\}
   \le\bar{u}<1 \]
and $\ddot{r}^\eps(t)=\dot{u}^\eps(t)\to 0$ as $t\to\infty$
for each $\eps\in ]0, \eps_2]$. Then for sufficient small $\eps$
the solutions have to lie on ${\cal I}_\eps$.
\end{proposition}
{\bf Proof\,:} Fix $\delta>0$. Since Theorem \ref{rigo} applies,
we find in the notation of Proposition \ref{kiba}
\[ |y(t)-h(x(t))| = |\dot{u}^\eps(t)+m(u^\eps(t))^{-1}\nabla 
V(r^\eps(t))|
   \le\delta/2 \]
for $t\ge t(\eps)$, with some $t(\eps)$. Thus the solution remains
$(\delta/2)$-close to ${\cal I}_0$ after time $t(\eps)$, and hence by
(\ref{1ord-darst}) also $\delta$-close to ${\cal I}_\eps$
for $\eps$ small. Since ${\cal I}_\eps$ is normally hyperbolic
(repulsive) at an $\eps$-independent rate and since $\delta>0$ was
arbitrary, this can only happen if the solution was already contained
in ${\cal I}_\eps$. {\hfill$\Box$}\bigskip

Corollary \ref{Q-bounds} provides a partial information on the long
time behavior of the solutions to (\ref{singul}) on the center manifold.
Roughly one can distinguish two classes. (i) (scattering): The
particle enters a domain where $-\nabla V = 0$ at $r_{1}$
with velocity $u_{\infty}$. If the straight line trajectory
$r_{1} + u_{\infty}t$, $t \geq 0$, is contained in this domain, then
the particle travels freely to infinity. Physically this is
a scattering trajectory. In this case $\lim_{t \to \infty} u(t)
= u_{\infty} \neq 0$, whereas the position has no limit.
(ii) (bounded motion): We assume that $|r(t)| \le const.$ and
that within this ball the critical points of $V$ form a discrete set.
Then by Corollary \ref{Q-bounds} and by continuity we have
$\lim_{t\to\infty} u(t) = 0$ and $\lim_{t\to\infty} r(t)=r_{\infty}$,
where $r_{\infty}$ is one of the critical points of $V$. If $r_{\infty}$
is a stable critical point, then the relaxation is exponentially
fast, as can be seen from linearization around the fixed point.
Clearly (i) and (ii) do not exhaust all possibilities.
The critical points of $V$ could lie on a sphere. If V is
confining, one would still expect convergence to a definite
$r_{\infty}$. Moreover, $V$ could vanish inside a ball. If $-\nabla V$
is pointing towards the ball, then close to each turning point
the particle looses energy. Thus $\lim_{t \to \infty}
u(t) = 0$, whereas the position has no limit. The potential could
decrease so slowly at infinity that no definite velocity is
approached. All these cases have to be studied separately.

Up to now we discussed  bounded potentials satisfying
$(U)$. In the introduction we claimed that
our results remain valid also for confining potentials satisfying
$(U')$. In this case, since $V$ is unbounded, we have no longer
${\cal K}_{\bar{v}}\subset \{H\le c_0\}$ for some $c_0\in\R$ as
in Theorem \ref{ZM-thm-U} above.
However, by energy conservation, one can derive the a priori bound
$\sup_{t\in\R} |q^\eps(t)|\le\bar{M}$ for solutions to the true system
(\ref{111}) on the macroscopic time scale.
Thus the motion is bounded also in the $q$-direction
and it suffices to build the center manifold for
the effective equation (\ref{ab-form}) over the bounded domain
\[ {\cal K}_{\bar{M}, \bar{v}}=\{r\in\R^3:
   |r|\le \bar{M}\}\times\{u\in\R^3: |u|\le\bar{v}\}\, , \]
enlarged to a suitable ${\cal K}_{\bar{M}+1, 1-\bar{\delta}}$
such that solutions starting over ${\cal K}_{\bar{M}, \bar{v}}$
stay away from the boundary of ${\cal K}_{\bar{M}+1, 1-\bar{\delta}}$
for $\eps>0$  sufficiently small. In this manner we obtain

\begin{theorem}\label{ZM-thm-Us}
Assume $(U')$ holds for the potential, and let
${\cal K}_{\bar{M}, \bar{\delta}}$ be defined as above.
Then there exists $\eps_1>0$ depending only on the initial data
such that for $\eps\in ]0, \eps_1]$ all solutions of (\ref{ab-form})
starting at points $(r, u, h_\eps(r, u))\in {\cal I}_\eps$,
$(r, u)\in {\cal K}_{\bar{M}, \bar{\delta}}$, exist globally.
Moreover, these solutions are uniformly bounded,
\begin{eqnarray}
   & \sup \{|r(t)|: t\in\R, \eps\in ]0, \eps_1]\}\le c(\bar{\delta}), &
   \nonumber \\
   & \sup \{|u(t)|: t\in\R, \eps\in ]0, \eps_1]\}
   \le 1-2\bar{\delta}<1,\quad\mbox{and}\quad &
   \nonumber \\
   & \sup\{|\dot{u}(t)|: t\in\R, \eps\in ]0, \eps_1]\}
   + \sup \{|\ddot{u}(t)|: t\in\R, \eps\in ]0, \eps_1]\}
   \le c(\bar{\delta}). & \nonumber\\ & & \label{Q-Us-bounds}
\end{eqnarray}
In addition,
\[ \dot{u}(t)\to 0,\quad\ddot{u}(t)\to 0,\quad\mbox{and}\quad
   \nabla V(r(t))\to 0\quad\mbox{as}\quad t\to\infty. \]
\end{theorem}
{\bf Proof\,:} The proof is similar to the one of Theorem \ref{ZM-thm-U}.
Concerning the boundedness, note that again for some $c_0\in\R$
and $\eps>0$ small,
\[ {\cal K}_{\bar{M}, \bar{\delta}}\subset\{H\le c_0\} \subset
   \{(r, u): G_\eps(r, u, h_\eps(r, u))\le c_0+c_a c_1\eps\}
   \subset\{H\le c_0+1\}. \]
Thus all solutions starting over ${\cal K}_{\bar{M}, \bar{\delta}}$
will remain on the manifolds over $\{H\le c_0+1\}$.
Since this set is independent of $\eps$ and compact by $(U')$,
the solutions must be uniformly bounded, because  $h_\eps$
is uniformly bounded. {\hfill$\Box$}\bigskip

On the center manifold the motion is governed by the (second order)
equation
\begin{equation}\label{h-form1}
   \dot{x} = f(x,h_{\eps}(x)).
\end{equation}
Since the existence of $h_{\eps}$ is established only abstractly,
Eq. (\ref{h-form1}) is somewhat implicit. From \cite[(2.9-1) \& 
Thm.~2.9]{saka}
we know that $h_{\eps}$ depends smoothly on $\eps$. Thus (\ref{h-form1})
can be expanded in $\eps$. Including the first Taylor term we pick 
up an error
of order $\eps^{2}$, which is of the same order as the error between 
the true
and the comparison dynamics on the center manifold. For consistency we
should stop then at this order. We make the ansatz
\[ h_\eps(r, u)=h_{0}(r, u)+\eps h_1(r, u)+h_{2, \eps} (r, u)
   \,,\quad |h_{2, \eps}(r, u)|\le c(\bar{\delta}) \eps^2 \]
for $(r, u)\in \R^3\times {\cal V}$. Then
\[ m(u)h_{0}(r,u) = -\nabla V(r), \]
and $h_1(r, u)$ is determined through
\[ D_x h_{0}(r, u)f(r, u, h_{0}(r, u))=D_y g(r, u, h_{0}(r, u), 0)
h_1(r, u)
   + D_\eps g(r, u, h(r, u), 0), \]
see \cite[(2.9-1) \& Thm.~2.9]{saka}. Computing the respective derivatives
one arrives at
\begin{eqnarray*}
    m(u)h_1(r, u) & = & a(u)\,\bigg[ -m(u)^{-1}\nabla^2 V(r)u
   \\ & & \hspace{2.8em} + {\bigg(\frac{d}{du}m(u)\bigg)}^{-1}\Big(
   \nabla V(r),
   m(u)^{-1}\nabla V(r)\Big)
   \\ & & \hspace{2.8em} +\,b(u, m(u)^{-1}\nabla V(r))\,\bigg]
\end{eqnarray*}
and the effective second order equation
\begin{equation}\label{second}
   \dot{r} = u\,, \quad m(u)\dot{u} = -\nabla V(r) + 
   \eps m(u) h_{1}(r,u)
\end{equation}
of the particle motion on the center manifold.


\section{Comparison of the true and the effective system}\label{main-sect}
\setcounter{equation}{0}

In this section we prove Theorem \ref{mainthm}.
Since we have $|u(t_1)|=|v^\eps(t_1)|\le\bar{v}$ by
(\ref{v-bounds}), Theorem \ref{ZM-thm-U}, resp.~Theorem
\ref{ZM-thm-Us}, implies that the solution trajectory
of the system with the modified functions $a(u)$, $m(u)$, and
$b(u, \dot{u})$ is indeed a solution trajectory to (\ref{ab-form}).
Recall that, by (\ref{ab-form-full}) and (\ref{ab-form}),
\begin{eqnarray}
   m(v^\eps)\dot{v}^\eps & = & -\nabla V(q^\eps) + \eps a(v^\eps)
   \ddot{v}^\eps + \eps b(v^\eps, \dot{v}^\eps) + \eps^2 f^\eps(t),
   \quad t\ge \eps t_1,
   \label{fulle-gl} \\
   m(u)\dot{u} & = & -\nabla V(r) + \eps a(u)\ddot{u}
   +\,\eps b(u, \dot{u}), \label{effe-gl}
\end{eqnarray}
with $|f^\eps(t)|\le C$. Using the bounds (\ref{v-bounds}) and
(\ref{Q-U-bounds}), resp.~(\ref{Q-Us-bounds}), we infer the weaker
estimate
\begin{eqnarray*}
   m(v^\eps)\dot{v}^\eps & = & -\nabla V(q^\eps) + {\cal O}(\eps),
   \quad t\ge t_1, \\
   m(u)\dot{u} & = & -\nabla V(r) + {\cal O}(\eps),
\end{eqnarray*}
which has been proved already in
\cite{KKS}. Hence (\ref{calv}) follows by the argument there.

To show (\ref{Hbound}) we compute  as in Lemma \ref{G-def},
using (\ref{fulle-gl}),
\begin{eqnarray*}
   \lefteqn{\frac{d}{dt} G_\eps(q^\eps(t), v^\eps(t), \dot{v}^\eps(t))}\\
   &=& \eps^2 f^\eps(t)v^\eps(t)\\
   & &  -\eps (\e^2/12\pi)
   \Big[\gamma(\dot{v}^\eps(t))^6 \dot{v}^\eps(t)^2
   +6\gamma(\dot{v}^\eps(t))^8 {(v^\eps(t)\cdot\dot{v}^\eps(t))}^2
   \Big]\,,\quad t\ge \eps t_1.
\end{eqnarray*}
Since $r(\eps t_1)=q^\eps(\eps t_1)$ and $u(\eps t_1)=
v^\eps(\eps t_1)$,
using the uniform bounds, we have for $t\ge \eps t_1$
\begin{eqnarray*}
   \lefteqn{|H(q^\eps(t), v^\eps(t))-H(r(t), u(t))|} \\ & \le &
   \eps\,\Big|(a(v^\eps(t))\dot{v}^\eps(t))\cdot v^\eps(t)-
   (a(u(t))\dot{u}(t))\cdot u(t)\Big| \\ & &
   + \int_{\eps t_1}^t ds\,\bigg[ \eps^2 |f^\eps(s) v^\eps(s)|
   +\eps (\e^2/12\pi)\bigg(\Big|\gamma(\dot{v}^\eps(s))^6 \dot{v}^\eps(s)^2
   -\gamma(\dot{u}(s))^6 \dot{u}^\eps(s)^2\Big|
   \\ & & \hspace{7em}
   +6\,\Big|\gamma(\dot{v}^\eps(s))^8 {(v^\eps(s)\cdot\dot{v}^\eps(s))}^2
   -\gamma(\dot{u}(s))^8 {(u(s)\cdot\dot{u}(s))}^2\Big|\bigg)\,\bigg]
   \\ & \le & C\eps\Big[|v^\eps(t)-u(t)|+|\dot{v}^\eps(t)-\dot{u}(t)|\Big]
   + C\eps^2 t \\ & & + C\eps\,\int_{\eps t_1}^t ds\,
   \Big[|v^\eps(s)-u(s)|+|\dot{v}^\eps(s)-\dot{u}(s)|\Big]
   \\ & \le & C\eps^2(1+t)\le C\eps^2,
\end{eqnarray*}
by (\ref{calv}) for $t={\cal O}(1)$. This concludes the proof of
Theorem \ref{mainthm}. {\hfill$\Box$}\bigskip

Finally we show that on a microscopic time scale our results
track the true trajectory with a higher precision,
cf.~(\ref{lis-pk}), Remark \ref{AW}(i).

\begin{proposition}\label{besser-bd} We have
\[ |q^\eps(t)-r(t)|\le c\eps^3 \quad\mbox{and}\quad
   |v^\eps(t)-u(t)|\le c\eps^2,\quad t={\cal O}(\eps), \]
i.e., (\ref{lis-pk}) holds.
\end{proposition}
{\bf Proof\,:} Define $\Psi (s)=\Big(\eps^{-1}q^\eps(\eps s)
-\eps^{-1}r(\eps s),v^\eps(\eps s)-u(\eps s), \eps\dot{v}^\eps (\eps s)
-\eps\dot{u}(\eps s)\Big)$ for $s\ge t_1$.
Then $\dot{\Psi}(s)=A\Psi(s)+\theta(s)$, where
$A=\left(\begin{array}{ccc}  0 & 1 & 0 \\ 0 & 0 & 1 \\ 0 & 0 & 0
\end{array}\right)$ and $\theta (s)=\eps^2(0, 0, \ddot{v}^\eps(\eps s)
-\ddot{u}(\eps s))$, whence
\begin{eqnarray*}
   |\theta (s)| & \le & c\eps\,\Big( |q^\eps(\eps s)-r(\eps s)| +
   |v^\eps(\eps s)-u(\eps s)| + |\dot{v}^\eps(\eps s)-\dot{u}(\eps s)| +
   \eps^2\Big) \\ & \le & c(|\Psi(s)|+\eps^3),\quad s\ge t_1,
\end{eqnarray*}
by (\ref{fulle-gl}), (\ref{effe-gl}), and the uniform bounds.
Therefore by the variation of constants formula and Gronwall's
inequality for $s\in [t_1, t_1+\tau]$, $|\Psi(s)|\le c(\tau)(|\Psi(s_1)|
+\eps^3)$. Consequently, $q^\eps(\eps t_1)=r(\eps t_1)$ and
$v^\eps(\eps t_1)=u(\eps t_1)$ yields
\[ \eps^{-1}|q^\eps(t)-r(t)| + |v^\eps(t)-u(t)| \le
   c(\tau)\Big(\eps |\dot{v}^\eps(\eps t_1)-\dot{u}(\eps t_1)|
   + \eps^3\Big) \]
for $t\in [\eps t_1, \eps t_1+\eps\tau]$. By (\ref{fulle-gl})
and (\ref{effe-gl}), $|\dot{v}^\eps(\eps t_1)-\dot{u}(\eps t_1)|\le c\eps$,
so that (\ref{lis-pk}) follows. {\hfill$\Box$}\bigskip


\section{Adiabatic limit of the fields}\label{feld-sect}
\setcounter{equation}{0}

We prove Theorem \ref{feld-thm}. Let $U(t)$ again denote
the fundamental solution of the wave equation
in $D^{1, 2}(\R^3)\oplus L^2(\R^3)$. We set
$Z(x,t)=(\phi(x, t), \pi(x, t))$ as well as
$\bar{\rho}_{\eps} = (0, \rho_{\eps})$. Then the inhomogeneous
wave equation in (\ref{111}) is solved as
\[ Z(x,t) = [U(t)Z(\cdot,0)](x)-\sqrt{\eps}\,\int_0^t ds\,[U(t-s)
   \bar{\rho}_\eps(\cdot-q^\eps(s))](x). \]
Since
\[ Z(x,0) =  - \sqrt{\eps}\,\int_{-\infty}^0 ds\,[U(s)
   \bar{\rho}_\eps (\cdot-q^0-v^0 s)](x), \]
cf.~Lemma \ref{regh}, we have for $t>0$
\[ Z(x,t) = - \sqrt{\eps}\,\int_{-\infty}^t ds\,[U(t-s)
   \bar{\rho}_\eps (\cdot - q^\eps(s))](x), \]
where we extended the position to negative times $t\le 0$
by $q^\eps(t)=q^0+v^0 t$. Thus by the solution formula
for the wave equation
\begin{equation}\label{knor}
   \frac{1}{\sqrt{\eps}}\,
   \phi(x, t) = -\int\frac{d^3 y}{4\pi|x-y|}\,
   \rho_\eps(y-q^\eps(t-|x-y|))
\end{equation}
and $\pi(x,t)=\dot\phi(x,t)$. For $\eps \to 0$, $q^\eps (t) \to
r(t)$, cf.~(\ref{li}), with $r(t)$ extended to negative times by
$r(t)= q^0+v^0 t$. Moreover, $\rho_{\eps}(x)=\eps^{-3}\rho(\eps^{-1}x)
\to\e\delta_0$ in the sense of distributions. Hence the
transformation $z=y-q^\eps(t-|x-y|)$, $\det (dy/dz)
={[1-v^\eps(t-|x-y|)\cdot (x-y)/|x-y|]}^{-1}$, in (\ref{knor})
yields the pointwise convergence (\ref{phi-lim}),
except on the worldline of the particle, since the integrand
in (\ref{knor}) is singular at $y=x$, i.e.~for $x=r(t)$
which corresponds to $t_{{\rm ret}}=t$.

The analogous argument works for $\pi(x,t)$. In the limit $\eps\to 0$,
$\pi$ is dis\-continuous at the light cone $\{x: |x|=t\}$, which we
avoided due to our assumption. {\hfill$\Box$}\bigskip


\section{Radiated Energy}\label{dissi-sect}
\setcounter{equation}{0}

Let $E_{R, q^\eps(t)} (t+R)$ be the energy, particle
plus field, at time $t+R$ in a ball of radius $R$ centered at $q^\eps(t)$.
For $R>\eps R_\rho$ this energy changes as
\begin{eqnarray}
   \lefteqn{\frac{d}{dt}\,\Bigg(E_{R, q^\eps(t)}(t+R)\bigg)}
   \nonumber\\ & = & \frac{d}{dt}
   \bigg( {\cal H}_{mac}(t=0) - \frac{1}{2} \int_{\{|x-q^\eps(t)|>R\}}
   d^3 x \Big[|\pi(x,t+R)|^2 + |\nabla \phi(x,t+R)|^2\Big] \bigg)
   \nonumber\\
   & = & R^2\,\int_{|\omega|=1} d^2\omega\,\pi(q^\eps(t)+R\omega, 
   t+R)\,
   \omega\cdot\nabla\phi (q^\eps(t)+R\omega, t+R) \nonumber\\
   & & + \frac{R^2}{2}\,\int_{|\omega|=1} d\omega\,(\omega\cdot
   v^\eps(t))
   \Big[ |\pi(q^\eps(t)+R\omega, t+R)|^2 \nonumber\\ & & \hspace{11em} +
   |\nabla\phi (q^\eps(t)+R\omega, t+R)|^2\Big], \label{rad}
\end{eqnarray}
where we used that the total energy is conserved.

$E_R$ changes because there is energy flowing
back and forth between particle and field,
and because energy is lost irreversibly to infinity. To separate 
both contributions
we take the limit $R\to\infty$. Using (\ref{knor}) and the relation 
$t+R-|q^\eps(t)+R\omega-y|
=t+\omega\cdot(y-q^\eps(t))+{\cal O}(1/R)$ for bounded $|y|$, we 
arrive at
\begin{eqnarray*}
   \lefteqn{I^\eps(t) = \lim_{R\to\infty}\,\frac{d}{dt}
   \bigg(E_{R, q^\eps(t)}(t+R)\bigg)}
   \\ & = & -\eps (4\pi)^{-2}\int_{|\omega|=1} d^2\omega\,
   (1-\omega\cdot v^\eps(t))\,
   \bigg[\int d^3 y\,\rho_\eps(y-q^\eps(t+\omega\cdot [y-q^\eps(t)]))
   \\ & & \hspace{15.5em} \times
   \frac{\omega\cdot\dot{v}^{\eps}(t+\omega\cdot [y-q^\eps(t)])}
   {(1-\omega\cdot v^\eps(t+\omega\cdot [y-q^\eps(t)]))^2}\bigg]^2
\end{eqnarray*}
cf.~\cite[Sec.~3]{KSK} for details on a similar calculation. In fact,
there the ball of radius $R$ was centered at the origin and the second
summand in (\ref{rad}) is absent. To let $\eps\to 0$, we again transform
to $z=y-q^\eps(t+\omega\cdot [y-q^\eps(t)])$, $\det (dy/dz)
=[1-\omega\cdot v^\eps(t+\omega\cdot [y-q^\eps(t)])]^{-1}$, use
$\rho_\eps(x)\to\e\delta_0$ in the sense of distributions,
and insert the identity $y=q^\eps(t)$ for $z=0$ to obtain
\begin{eqnarray*}
   \lim_{\eps\to 0} \eps^{-1}\,I^\eps(t) & = & -\e^2 (4\pi)^{-2}\,
   \int_{|\omega|=1} d^2\omega\,(1-\omega\cdot u(t))^{-5}\,
   (\omega\cdot\dot u(t))^2 \\
   & = & -(\e^2/12\pi)\,[6\gamma^8 (u(t)\cdot\dot u(t))^2
   +\gamma^6\dot u(t)^2],
\end{eqnarray*}
in agreement with (\ref{lya}).

Alternatively, we could first take the limit $\eps\to 0$ in
(\ref{rad}). Using Theorem 2.4 we find, with
$(\overline\phi, \overline\pi)$ denoting the limit fields
from (\ref{phi-lim}), (\ref{pi-lim}),
\begin{eqnarray*}
   I_R(t) & = & \lim_{\eps\to 0}\,\eps^{-1}\,\frac{d}{dt}
   \bigg(E_{R, q^\eps(t)}(t+R)\bigg)
   \\ & = & R^2 \int_{|\omega|=1} d^2\omega\,\overline\pi(r(t)+
   R\omega,t+R)
   \,\omega\cdot\nabla\overline\phi (r(t)+R\omega, t+R) \\
   & & + \frac{R^2}{2} \int_{|\omega|=1} d^2\omega\,(\omega\cdot u(t))\,
   \Big[ |\overline\pi(r(t) + R\omega, t+R)|^2 \\ & & \hspace{11em}
   + |\nabla\overline\phi (r(t) + R\omega, t+R)|^2\Big].
\end{eqnarray*}
Since both $\overline\pi$ and $\nabla\overline\phi$ have one term 
proportional
to $R^{-1}$ and other contributions of order $R^{-2}$, in the
limit $R\to\infty$ only the product of the two leading terms survives
and it follows that
\[ \lim_{R\to\infty}\,I_R(t)= -\e^2 (4\pi)^{-2}\int_{|\omega|=1} 
d^2\omega
   \,(1-\omega\cdot u(t))^{-5}\,(\omega\cdot\dot u(t))^2,  \]
as before. We note that the radiated energy is of order $\eps$ and it
therefore suffices to use the effective dynamics to order one, i.e.,
ignoring the radiation reaction.


\section{Appendix: Proof of Lemma \ref{q4-bound}}\label{append}
\setcounter{equation}{0}

In this appendix we prove Lemma \ref{q4-bound}. Since we need to
use some identities from \cite{KKS}, we switch back to the
original time scale of (\ref{system}). Hence we have to show \medskip

\noindent {\bf Lemma \ref{q4-bound}}\,\,{\em For solutions of 
(\ref{system})
with initial values satisfying (\ref{ini-cond}), i.e., starting
on the soliton manifold, and for ${|\rho|}_{L^2}$ sufficiently small 
we have}
\[ \sup_{t\in\R}|\stackrel{...}{v}(t)|\le C\eps^3\,. \]
{\em The constant $C$ and the bound on ${|\rho|}_{L^2}$ depend
only on the data.} \bigskip

\noindent
{\bf Proof\,:} From \cite[Lemma 2.2 and Prop.~4.1]{KKS} we already
know the bounds
\begin{eqnarray}\label{q23-bounds}
   & \sup_{t\in\R}|v(t)|\le \bar{v}<1\,,\quad
   \sup_{t\in\R}|\dot{v}(t)|+\sup_{t\in\R}|\dot{p}(t)|
   \le C\eps & \nonumber\\ & \mbox{and}\quad
   \sup_{t\in\R}|\ddot{v}(t)|+\sup_{t\in\R}|\ddot{p}(t)|\le C\eps^2, &
\end{eqnarray}
for ${|\rho|}_{L^2}$ sufficiently small. The constants $\bar{v}$ and $C$
appearing in (\ref{q23-bounds}) do not depend on the particular
solution, but only on bounds for the initial values.

Denote
\[ Z(x, t)=\left(\begin{array}{c} \varphi(x, t) \\ \psi(x, t)
   \end{array}\right)=\left(\begin{array}{c}
   \phi(x, t)-\phi_{v(t)}(x-q(t)) \\ \pi(x, t)-\pi_{v(t)}(x-q(t))
   \end{array}\right)\,. \]
Then, cf.~\cite{KKS}, with $L(t)\phi=\nabla\phi\cdot v(t)+\dot{\phi}$,
\begin{eqnarray*}
   \ddot{p}(t) & = & -\eps^2 \nabla^2 V(\eps q(t))\cdot v(t)
   + \int d^3x\,(L(t)\varphi)(x+q(t), t)\nabla\rho(x) \\
   & =: & -\eps^2 \nabla^2 V(\eps q(t))\cdot v(t)+M(t).
\end{eqnarray*}   
Therefore
\begin{equation}\label{pdrei}
   \stackrel{...}{p}(t)=-\eps^2\nabla^2 V(\eps q(t))\cdot\dot{v}(t)
   -\eps^3\nabla^3 V(\eps q(t))(v(t), v(t)) + \dot{M}(t)\,.
\end{equation}
Below we will show
\begin{lemma}\label{dotM-estim} The estimate
   \[ |\dot{M}(t)|\le C\,\bigg(\eps^3+{|\rho|}_{L^2}\,\int_0^t\,
   \frac{|\stackrel{...}{v}(s)|}{1+(t-s)^2}\,ds\bigg) \]
holds.
\end{lemma}
Then according to (\ref{pdrei}), (\ref{q23-bounds}), and assumption
$(U)$ on the potential,
\begin{equation}\label{bona}
   |\stackrel{...}{p}(t)| \le C\,\bigg(\eps^3+{|\rho|}_{L^2}\,\int_0^t\,
   \frac{|\stackrel{...}{v}(s)|}{1+(t-s)^2}\,ds\bigg)\,.
\end{equation}
Since
\begin{eqnarray*}
   \lefteqn{|\stackrel{...}{v}|} \\ & = &
   \bigg|\,\frac{d}{dp}\bigg(\frac{p}{\sqrt{1+p^2}}\bigg)
   \stackrel{...}{p} + 3\,
   \frac{d^2}{dp^2}\bigg(\frac{p}{\sqrt{1+p^2}}\bigg)
   (\dot{p}, \ddot{p}) +
   \frac{d^3}{dp^3}\bigg(\frac{p}{\sqrt{1+p^2}}\bigg)
   (\dot{p}, \dot{p}, \dot{p})\,\bigg| \\
   & \le & C (|\stackrel{...}{p}|+\eps^3),
\end{eqnarray*}   
the claim of Lemma \ref{q4-bound} obtains from (\ref{bona})
by taking ${|\rho|}_{L^2}$ small enough. {\hfill$\Box$}\bigskip

Thus it remains to give the \smallskip

\noindent
{\bf Proof of Lemma \ref{dotM-estim}\,:} First note
\[ \dot{M}(t)=\int d^3x\,{\Big\langle ({\cal L}(t)Z(\cdot, t))(x),\,
   \nabla\rho_\ast(x-q(t))\Big\rangle}_{\R^2}\,,\quad
   \rho_\ast(x)=(\rho(x), 0), \]
where ${\cal L}(t)Z=\nabla Z\cdot\dot{v}(t)+(\nabla^2 Z)
(v(t), v(t))+2\nabla\dot{Z}\cdot v(t)+\ddot{Z}$. Because
$\dot{Z}=AZ-B$, with
\begin{equation}\label{AB-def}
   A(\phi, \pi)=(\pi, \Delta\phi)\quad\mbox{and}\quad
   B(x, t)=\left(\begin{array}{c} \nabla_v\phi_{v(t)}(x-q(t))
   \cdot\dot{v}(t)
   \\ \nabla_v\pi_{v(t)}(x-q(t))\cdot\dot{v}(t)\end{array}\right)\,,
\end{equation}
we obtain
\[ \frac{d}{dt}({\cal L}(t)Z)=A({\cal L}(t)Z)-{\cal L}(t)B
   + 2\,\big[(\nabla^2 Z)(v, \dot{v})+\nabla\dot{Z}\cdot\dot{v}\big]
   + \nabla Z\cdot\ddot{v}\,. \]
Let $U(t)$ again denote the group generated by the free wave equation on
$D^{1, 2}(\R^3)\oplus L^2(\R^3)$. Then
\begin{eqnarray*}
   \dot{M}(t) & = & {\Big\langle U(t)[{\cal L}(0)Z(\cdot, 0)],\,
   \nabla\rho_\ast(\cdot-q(t))\Big\rangle}_{L^2(\R^2)} \\ & & +
   \int_0^t ds\,\bigg[\,-{\Big\langle U(t-s)[{\cal L}(s)B(\cdot, s)],\,
   \nabla\rho_\ast(\cdot-q(t))\Big\rangle}_{L^2(\R^2)} \\ & & 
   \hspace{4.3em}
   +\,2\,\Big\langle U(t-s)[(\nabla^2 Z(\cdot, s))(v(s),
   \dot{v}(s))+\nabla\dot{Z}(\cdot, s)\cdot\dot{v}(s)],\,
   \\ & & \hspace{6.5em}
   \nabla\rho_\ast(\cdot-q(t))\Big\rangle_{L^2(\R^2)} \\ & &
   \hspace{4.3em} +\,{\Big\langle U(t-s)[\nabla Z(\cdot, s)
   \cdot\ddot{v}(s)],\,
   \nabla\rho_\ast(\cdot-q(t))\Big\rangle}_{L^2(\R^2)}\,\bigg]
   \\ & =: & T_0 + T_1 + T_2 + T_3\,.
\end{eqnarray*}
We estimate each term $T_j$ separately, keeping all parts which
do contain only initial values. Note that here according to
(\ref{ini-cond}) we have $Z(x, 0)=0$, so all these terms vanish.
Nevertheless, we wanted to derive the general form of the estimate;
see Remark \ref{AW}(iii). \smallskip

\underline{Estimate of $T_3$:} Since $\dot{Z}=AZ-B$, we find
\begin{equation}\label{Z-repr}
   Z(t)=U(t)Z(0)-\int_0^t ds\,U(t-s)B(\cdot, s),
\end{equation}
and hence
\begin{eqnarray*}
   T_3 & = & \int_0^t ds\,
   {\Big\langle U(t)[\nabla Z(\cdot, 0)\cdot\ddot{v}(s)],\,
   \nabla\rho_\ast(\cdot-q(t))\Big\rangle}_{L^2(\R^2)}
   \\ & & -
   \int_0^t ds\,\int_0^s d\tau\,
   {\Big\langle U(t-\tau)[\nabla B(\cdot, \tau)\cdot\ddot{v}(s)],\,
   \nabla\rho_\ast(\cdot-q(t))\Big\rangle}_{L^2(\R^2)} \\
   & =: & T_{3, 0} + T_{3, 1}\,.
\end{eqnarray*}
Then Lemma \ref{abfall} below and (\ref{q23-bounds}) imply through
integration by parts in the $d^3x$-integral,
\[ T_{3, 1}\le C\eps^2\,\int_0^t ds\,\int_0^s d\tau\,
   \frac{\eps}{1+(t-\tau)^3} \le C\eps^3\,. \]
\smallskip

\underline{Estimate of $T_0$:} This term is determined solely
through the data. \smallskip

\underline{Estimate of $T_1$:} If we calculate the form of ${\cal L}(t)B
=\nabla B\cdot\dot{v}+(\nabla^2 B)(v, v)+2\nabla\dot{B}\cdot
v+\ddot{B}$ explicitly from (\ref{AB-def}), fortunately many terms cancel,
and we find with $\displaystyle\Phi_v=\left(\begin{array}{c} \phi_v
\\ \pi_v\end{array}\right)$,
\[ {\cal L}(t)B=
   \nabla_v\Phi_v(x-q)\cdot\stackrel{...}{v}
   +3\nabla_v^2\Phi_v(x-q)(\dot{v}, \ddot{v})
   +\nabla^3_v\Phi_v(x-q)(\dot{v}, \dot{v}, \dot{v}) \,. \]
Now we may argue analogously to the estimate of $T_3$ and Lemma 
\ref{abfall}
to obtain with
\[ \left(\begin{array}{c} \tilde{\phi}(x) \\ \tilde{\pi}(x)\end{array}
   \right) = [U(t-s){\cal L}(s) B(\cdot, s)](x)\,, \]
the estimate
\begin{equation}\label{lauch}
   |\nabla\tilde{\phi}(x+q(t))|
   \le \frac{C}{1+(t-s)^2}\,\big(\eps^3+|\stackrel{...}{v}(s)|\big)\,,
   \quad |x|\le R_\rho\,,\quad t\ge s\,.
\end{equation}
Here we have used (\ref{q23-bounds}) and some of the estimates
\begin{eqnarray}\label{vers}
   |\nabla\nabla_v\phi_v(x)| + |\nabla\nabla_v^2\phi_v(x)|
   + |\nabla\nabla_v^3\phi_v(x)| & \le & C{(1+|x|)}^{-2}\,,\nonumber \\
   |\nabla^2\nabla_v\phi_v(x)| + |\nabla^2\nabla_v^2\phi_v(x)|
   + |\nabla^2\nabla_v^3\phi_v(x)| & \le & C{(1+|x|)}^{-3}\,,\nonumber \\
   |\nabla^3\nabla_v\phi_v(x)| + |\nabla^3\nabla_v^2\phi_v(x)|
   + |\nabla^3\nabla_v^3\phi_v(x)| & \le & C{(1+|x|)}^{-4}\,,\nonumber \\
   |\nabla^4\nabla_v\phi_v(x)| + |\nabla^4\nabla_v^2\phi_v(x)|
   + |\nabla^4\nabla_v^3\phi_v(x)| & \le & C{(1+|x|)}^{-5}\,,\nonumber \\
   |\nabla\nabla_v\pi_v(x)| + |\nabla\nabla_v^2\pi_v(x)|
   + |\nabla\nabla_v^3\pi_v(x)| & \le & C{(1+|x|)}^{-3}\,,\nonumber \\
   |\nabla^2\nabla_v\pi_v(x)| + |\nabla^2\nabla_v^2\pi_v(x)|
   + |\nabla^2\nabla_v^3\pi_v(x)| & \le & C{(1+|x|)}^{-4}\,,\nonumber \\
   |\nabla^3\nabla_v\pi_v(x)| + |\nabla^3\nabla_v^2\pi_v(x)|
   + |\nabla^3\nabla_v^3\pi_v(x)| & \le & C{(1+|x|)}^{-5},
\end{eqnarray}
for $x\in\R^3$ and $|v|\le\bar{v}$. From (\ref{lauch}) we conclude
\[ T_1=-\int_0^t ds\,\int_{|x|\le R_\rho}
   d^3x\,\nabla\tilde{\phi}(x+q(t))\rho(x)
   \le C\,\bigg(\eps^3+{|\rho|}_{L^2}\,\int_0^t\,
   \frac{|\stackrel{...}{v}(s)|}{1+(t-s)^2}\,ds\bigg). \]
\smallskip

\underline{Estimate of $T_2$:} Let $P(t)Z
=\nabla^2 Z (\cdot, t)v(t)+\nabla\dot{Z}(\cdot, t)$. Then
\[ \frac{d}{dt}(P(t)Z) = P(t)\dot{Z}+(\nabla^2 Z)\dot{v}
   = A(P(t)Z)-P(t)B+(\nabla^2 Z)\dot{v}\,. \]
Therefore by definition of $T_2$,
\begin{eqnarray*}
    T_2 & = & 2\,\int_0^t ds\,{\Big\langle U(t)
    [(P(0)Z(\cdot, 0))\cdot\dot{v}(s)],\,
    \nabla\rho_\ast(\cdot-q(t))\Big\rangle}_{L^2(\R^2)}
    \\ & & + 2\,\int_0^t ds\,\int_0^s d\tau\,\Big\langle U(t-\tau)
    \Big[ -P(\tau)B(\cdot, \tau)+(\nabla^2 Z(\cdot, \tau))\dot{v}(\tau)
    \Big]\cdot\dot{v}(s), \\ & & \hspace{7.4em}
    \,\nabla\rho_\ast(\cdot-q(t))\Big\rangle_{L^2(\R^2)}
    \\ & =: & T_{2, 0} + T_{2, 1} + T_{2, 2}\,.
\end{eqnarray*}
To estimate $T_{2, 1}$, observe
\[ P(t)B=\nabla\nabla_v\Phi_v(x-q)\cdot\ddot{v}+
   \nabla\nabla_v^2\Phi_v(x-q)(\dot{v}, \dot{v})\,. \]
Hence we may argue as before to find $|T_{2, 1}|\le C\eps^3$.
In order to bound $T_{2, 2}$, similarly to the estimate of
$T_3$ we again use (\ref{Z-repr}) to get
\begin{eqnarray*}
   T_{2, 2} \\ & = & 2\,\int_0^t ds\,\int_0^s d\tau\,{\Big\langle U(t)
   [\nabla^2 Z(0)(\dot{v}(\tau), \dot{v}(s))],\,
   \nabla\rho_\ast(\cdot-q(t))\Big\rangle}_{L^2(\R^2)} \\
   & & - 2\,\int_0^t ds\,\int_0^s d\tau\,\int_0^\tau d\sigma\,
   \Big\langle U(t-\sigma)
   [\nabla^2 B(\cdot, \sigma)(\dot{v}(\tau), \dot{v}(s))],
   \\ & & \hspace{10.2em}
    \,\nabla\rho_\ast(\cdot-q(t))\Big\rangle_{L^2(\R^2)}
   \\ & =: & T_{2, 2, 0} + T_{2, 2, 1}\,.
\end{eqnarray*}
By (\ref{vers}) and the argument of Lemma \ref{abfall} then
\[ T_{2, 2, 1} \le \int_0^t ds\,\int_0^s d\tau\,\int_0^\tau d\sigma\,
   \frac{C\eps^3}{1+(t-\sigma)^4} \le C\eps^3\,. \]
\smallskip

Summarizing all above estimates for $T_0$--$T_3$, we hence arrive at
\begin{eqnarray}\label{data-null}
   |\dot{M}(t)| & \le & C\,\bigg(\eps^3+{|\rho|}_{L^2}\,\int_0^t\,
   \frac{|\stackrel{...}{v}(s)|}{1+(t-s)^2}\,ds\bigg)
   \nonumber \\ & & +\,{\Big\langle U(t)[{\cal L}(0)Z(\cdot, 0)],\,
   \nabla\rho_\ast(\cdot-q(t))\Big\rangle}_{L^2(\R^2)} \nonumber \\ & &
   +\,2\,\int_0^t ds\,{\Big\langle U(t)
    [(P(0)Z(\cdot, 0))\cdot\dot{v}(s)],\,
    \nabla\rho_\ast(\cdot-q(t))\Big\rangle}_{L^2(\R^2)} \nonumber \\ & &
   +\,2\,\int_0^t ds\,\int_0^s d\tau\,{\Big\langle U(t)
   [\nabla^2 Z(0)(\dot{v}(\tau), \dot{v}(s))],\,
   \nabla\rho_\ast(\cdot-q(t))\Big\rangle}_{L^2(\R^2)} \nonumber \\ & &
   +\,\int_0^t ds\,
   {\Big\langle U(t)[\nabla Z(\cdot, 0)\cdot\ddot{v}(s)],\,
   \nabla\rho_\ast(\cdot-q(t))\Big\rangle}_{L^2(\R^2)}\,.
\end{eqnarray}
Concerning the terms that contain data, these vanish here since
$Z(x, 0)=0$ as a consequence of (\ref{ini-cond}).
This completes the proof of Lemma \ref{dotM-estim}. {\hfill$\Box$}\bigskip

In case of solutions starting not on, but close, to the soliton manifold
as discussed in Remark \ref{AW}(iii), conditions on the data have to 
be imposed
to ensure the last four terms in (\ref{data-null}) can also be
estimated by $C\eps^3$. In \cite[Thm.~2.6]{KKS} and Section 4 of that paper
details are carried out for derivatives of one order less.

Above we used the following lemma.

\begin{lemma}\label{abfall} The estimate
\begin{equation}\label{cj}
   {\|\nabla [U(t-\tau)\nabla B(\cdot, \tau)](\cdot+q(t))\|}_{R_\rho}
   \le C\,\frac{\eps}{1+(t-\tau)^3}\,,\quad t\ge\tau\,,
\end{equation}
holds.
\end{lemma}
{\bf Proof\,:} Such estimates have already been used in \cite{KKS}, but we
nevertheless include some details of the argument. Let
\[ \left(\begin{array}{c} \tilde{\phi}(x) \\ \tilde{\pi}(x)\end{array}
   \right) = [U(t-\tau)\nabla B(\cdot, \tau)](x) \]
for fixed $t, \tau$. By Kirchhoff's formula for the solution
to the wave equation and by (\ref{AB-def}),
\begin{eqnarray}
   \lefteqn{\nabla\tilde{\phi}(x+q(t))} \nonumber\\
   & = & \frac{1}{4\pi (t-\tau)^2}\,
   \int_{|y-x-q(t)|=(t-\tau)} d^2y\,\Big[ (t-\tau)
   \nabla^2\nabla_v\pi_{v(\tau)}(y-q(\tau))\cdot\dot{v}(\tau)
   \nonumber \\ & & \hspace{11em}
   + \nabla^2\nabla_v\phi_{v(\tau)}(y-q(\tau))\cdot\dot{v}(\tau)
   \nonumber \\[0.5ex] & & \hspace{11em}
   + \nabla^3\nabla_v\phi_{v(\tau)}(y-q(\tau))(\dot{v}(\tau),
   y-x-q(t))\,\Big]. \nonumber \\& & \label{kimb}
\end{eqnarray}
Now $|x|\le R_\rho$ and $|y-x-q(t)|=(t-\tau)$ yields
$|y-q(\tau)|\ge (t-\tau)-\bar{v}(t-\tau)-R_\rho
=(1-\bar{v})(t-\tau)-R_\rho$ by (\ref{q23-bounds}). As a consequence of
(\ref{vers}), hence (\ref{cj}) follows from (\ref{kimb}).
{\hfill$\Box$}\bigskip

\noindent {\bf Acknowledgement:} We thank A.~Komech for
useful discussions.

\end{document}